\documentclass[twocolumn,superscriptaddress,prb,amsmath,amstex,amssymb,citeautoscript,longbibliography]{revtex4-1}
\pdfoutput=1

\usepackage[toc,page]{appendix}
\usepackage{rotating}
\usepackage{natbib}
\usepackage[english]{babel}
\usepackage{letltxmacro}
\usepackage{latexsym}
\usepackage[utf8]{inputenc}
\LetLtxMacro{\ORIGselectlanguage}{\selectlanguage}
\makeatletter
\DeclareRobustCommand{\selectlanguage}[1]{%
  \@ifundefined{alias@\string#1}
    {\ORIGselectlanguage{#1}}
    {\begingroup\edef\x{\endgroup
       \noexpand\ORIGselectlanguage{\@nameuse{alias@#1}}}\x}%
}
\newcommand{\definelanguagealias}[2]{%
  \@namedef{alias@#1}{#2}%
}
\makeatother
\definelanguagealias{en}{english}
\definelanguagealias{English}{english}
\usepackage{graphicx}
\usepackage{amsmath}
\usepackage{amsfonts}
\usepackage{amsthm}   % Theorem Formatting
\usepackage{amssymb}
\usepackage{bm}
\usepackage{color}
\usepackage[percent]{overpic}
\usepackage{soul} 
\usepackage{amssymb}
\usepackage{wasysym}
\usepackage{dsfont}
\usepackage{float}

\usepackage{hyperref}
\hypersetup{
    bookmarks=false,         % show bookmarks bar?
    unicode=false,          % non-Latin characters in Acrobat�s bookmarks
    pdftoolbar=false,        % show Acrobat�s toolbar?
    pdfmenubar=true,        % show Acrobat�s menu?
    pdffitwindow=false,     % window fit to page when opened
    pdfstartview={FitH},    % fits the width of the page to the window
    pdftitle={},    % title
    pdfauthor={Authors},     % author
    pdfsubject={},   % subject of the document
    pdfcreator={},   % creator of the document
    pdfproducer={}, % producer of the document
    pdfkeywords={quantum many-body scars} {lie algebra} {non-equilibrium dynamics}, % list of keywords
    pdfnewwindow=true,      % links in new window
    colorlinks=true,       % false: boxed links; true: colored links
    linkcolor=black,          % color of internal links (change box color with linkbordercolor)
    citecolor=blue,        % color of links to bibliography
    filecolor=magenta,      % color of file links
    urlcolor=blue           % color of external links
}

\newcommand{\be}{\begin{equation}}
\newcommand{\ee}{\end{equation}}
\newcommand{\bea}{\begin{eqnarray}}
\newcommand{\eea}{\end{eqnarray}}

\usepackage{graphicx}
\usepackage[colorinlistoftodos]{todonotes}
\usepackage{verbatim}

\newcommand{\ket}[1]{\mbox{$| #1 \rangle$}}

\newcommand{\Tr}{\mathrm{Tr}}

\usepackage{soul}
\usepackage[normalem]{ulem}

\newtheorem*{lemma*}{Lemma}

\begin{document}

\title{Quantum scars as embeddings of weakly ``broken'' Lie algebra representations}

\begin{abstract}
{
We present an interpretation of scar states and quantum revivals as weakly ``broken" representations of Lie algebras spanned by a subset of eigenstates of a many-body quantum system. We show that the PXP model, describing strongly-interacting Rydberg atoms, supports a ``loose" embedding of multiple $\mathrm{su(2)}$ Lie algebras corresponding to distinct families of scarred eigenstates. Moreover, we demonstrate that these embeddings can be made progressively more accurate via an iterative process which results in optimal perturbations that stabilize revivals from arbitrary charge density wave product states, $|\mathbb{Z}_n\rangle$, including ones that show no revivals in the unperturbed PXP model. We discuss the relation between the loose embeddings of Lie algebras present in the PXP model and recent exact constructions of scarred states in related models. 
}
\end{abstract}

\author{Kieran Bull, Jean-Yves Desaules, and Zlatko Papi\'c}
\affiliation{School of Physics and Astronomy, University of Leeds, Leeds LS2 9JT, United Kingdom}
\maketitle

\section{Introduction}
Isolated quantum systems are expected to approach thermal equilibrium after sufficiently long times and much of current research focuses on understanding the conditions for this to happen as well as the details of the process of thermalization~\cite{Gogolin2016}. From this point of view, quantum revival -- a wave function periodically returning to its value at time $t=0$~\cite{Bocchieri1957,Percival1961} -- is a well-known  counterexample of non-thermalizing dynamics that has played an important role since the early days of quantum physics. Experimentally, such recurrent behaviour has been observed in small or weakly-interacting quantum systems, for example the Jaynes-Cummings model describing a two-level atom interacting with a resonant monochromatic field~\cite{Eberly1980}, a micromaser cavity with rubidium atom~\cite{Rempe1987}, in a Rydberg electron wave packet~\cite{Yeazell1990},  vibrational wave packets in $\mathrm{Na}_2$~\cite{Baumert1992}, infinite square well potentials and various types of billiards~\cite{Robinett2002, Aronstein1997, Dubois2017}, cold atoms~\cite{Brune1996,Greiner2002,Will2010}, and more recently larger systems of one-dimensional superfluids~\cite{Schweigler2017,Rauer2018}. The ability to engineer recurrent behavior in more complex quantum many-body systems is an important task because this allows one to study their long-term coherent evolution beyond the initial relaxation, while on the other hand, it also provides insight into the emergence of statistical ensembles in closed quantum systems that evolve according to the Schr\"odinger unitary evolution.

Intuitively, the conditions for observing \emph{many-body} wave function revivals in a strongly-interacting quantum system are expected to be very stringent due to the exponentially large size of the Hilbert space. It was thus surprising when recent experiments on strongly-interacting one-dimensional chains of Rydberg atoms~\cite{Schauss2012,Labuhn2016} observed revivals of local observables when the chain was quenched~\cite{CalabreseQuench} from an initial N\'eel state of atoms~\cite{Bernien2017},  $|\psi(0)\rangle = |\mathbb{Z}_2\rangle \equiv | 0101\ldots\rangle$, where $0$ denotes an atom in the ground state and $1$ in the excited (Rydberg) state. This observation was surprising as the N\'eel state effectively forms an ``infinite-temperature" ensemble for this system, for which equilibration is expected to occur very fast according to the Eigenstate Thermalization Hypothesis (ETH)~\cite{DeutschETH, SrednickiETH}. The observed revivals were thus in apparent disagreement with the na\"ive expectations based on the ETH. Moreover,
the revivals from the N\'eel initial state have also been seen in numerical simulations of an idealized model believed to describe the Rydberg atom chain~\cite{Sun2008,LesanovskyDynamics,Olmos2012,Turner2017,wenwei18TDVPscar}.
This model is known as the ``PXP" model~\cite{Bernien2017}, and it has the form of a one-dimensional spin-1/2 chain with a kinetically-constrained spin flip term that results from removing all nearest-neighbor pairs of atoms that are simultaneously excited into  the Rydberg states (see Sec.~\ref{sec:pxp} for more details on the model). It has been understood that the key to revivals in the Rydberg atom chain are the special eigenstates -- ``quantum many-body scars"~\cite{Turner2017,lin2018exact} --  whose non-thermal properties cause a violation of the strong ETH~\cite{dAlessio2016, Gogolin2016, ShiraishiMori}. Such atypical eigenstates have previously been rigorously constructed in the non-integrable Affleck-Kennedy-Lieb-Tasaki (AKLT) model~\cite{Bernevig2017,BernevigEnt}. While the collection of models that feature scarred-like eigenstates has recently expanded~\cite{Calabrese16, Konik1, Konik2, Vafek, IadecolaZnidaric, NeupertScars, Haldar2019,Moudgalya2019,Pretko2019,Khemani2019,Sala2019,Khemani2019_2}, a smaller subset of such models have been demonstrated to display revivals from easily preparable initial states~\cite{Bull2019,Michailidis2019,Buca2019_2}. Thus, the connection between revivals and the presence of atypical eigenstates remains to be fully understood.
 
Revivals in the experimentally realized PXP model are relatively fragile. For example,  numerical simulations have shown that the revival of a wavefunction, quantified in terms of the return probability, $\vert \langle \psi(0) \vert \psi(t) \rangle \vert^2$~\cite{Gorin2006}, is at best $\sim70$\% of its initial value, and it undergoes a clear decay as a function of time~\cite{TurnerPRB}. While the imperfect PXP revivals are still remarkable given the exponentially large many-body Hilbert space, their decay poses a question of whether the PXP many-body scars could be a transient effect that disappears in the thermodynamic limit.  It was realized, however, that revivals can be significantly enhanced by slightly deforming the PXP model~\cite{Khemani2018}, with the fidelity revival reaching the value $\sim (1-10^{-6})$ in the largest systems available in numerics~\cite{Choi2018}, suggesting there could exist fine-tuned models that host ``perfect" many-body scars while their overall behavior, as witnessed by the energy level statistics~\cite{Choi2018}, remains thermalizing. 

Indeed, several non-integrable spin chain models have recently been shown to contain ``exact" scars and exhibit perfect wavefunction revivals when quenched from special initial states~\cite{Iadecola2019_2, Iadecola2019_3,Chattopadhyay,OnsagerScars}. 
Exact revivals in these models are a consequence of a dynamical symmetry of certain terms in the Hamiltonian (as we explain below in Sec.~\ref{section:exact_embedding}), such that scarred eigenstates are equidistant in energy. On the other hand, PXP is not the only model to exhibit decaying wavefunction revivals due to many-body quantum scars. This phenomenon has also been observed in models of fractional quantum Hall effect in a quasi one-dimensional limit~\cite{Moudgalya2019} and in a model of bosons with correlated hopping~\cite{bosonScars}. In each of these cases it was found that scar states are well approximated by Ritz vectors of a Krylov-like subspace generated by the action of some raising operator. In general, the energy variance of this subspace is non-zero; however, provided this subspace variance is small, the Hamiltonian takes the approximate block-diagonal form $H \approx H_{\mathrm{Krylov}} \bigoplus H_{\bot}$. This is reminiscent of the recently introduced notion of ``Krylov-restricted thermalization"~\cite{MoudgalyaKrylov}, whereby the Hilbert space fractures into closed Krylov subspaces in which exponentially large integrable and ergodic sectors can coexist alongside one another. While ``Krylov restricted thermalization" with exponentially large integrable sectors arises naturally in a model of interacting fermions~\cite{MoudgalyaKrylov}, it has also been demonstrated that one can embed a target integrable subspace of arbitrary size alongside ergodic subspaces in an interacting spin model~\cite{ShiraishiMori,NeupertScars}. We will refer to the latter approach as ``projector embedding".

In this paper we demonstrate how a ``loosely embedded" integrable subspace can give rise to many-body quantum scars and strong ETH violation, thus providing a general picture  of scarring in the PXP model that relates it  to other types of scarred models in the literature. Our embedding scheme is defined by considering Hamiltonians that consist of generators of a Lie algebra representation, but with slightly ``broken" commutation relations, resulting in the approximate block diagonal form $H \approx H_{\mathrm{int}} \bigoplus H_{\bot}$. Due to the ``broken" root struture of the Lie algebra, $H_{\mathrm{int}}$ is found to possess an approximate dynamical symmetry such that scar states are embedded throughout the spectrum with nearly equal energy spacing. This, along with the non-zero subspace variance, gives rise to decaying wavefunction revivals when the system is quenched from certain initial states.

Further, we introduce an iterative scheme to identify perturbations which correct the errors in the root structure of the Lie algebra representation. While the perturbations we find are generically long-range and have complicated forms, they serve to elucidate the connection between exact integrable subspaces, seen in either ``projector embedding" or ``Krylov-restricted thermalization", and loose embeddings such as in PXP model. Correcting the algebra causes the energy variance of the loosely embedded subspace to decrease, resulting in the Hamiltonian becoming increasingly block diagonal. In addition, an improving root structure within the embedded subspace results in scar states becoming more equidistant in energy, such that revivals are also enhanced. 

Specifically, our scheme allows to re-derive perturbations to the PXP model which have been shown to enhance revivals from the $\vert \mathbb{Z}_2 \rangle$ state~\cite{Khemani2018,Choi2018}. Nevertheless, in doing so, we also identify a missing set of perturbations which enhance the revivals further by several orders of magnitude compared to previous works~\cite{Khemani2018,Choi2018}. Moreover, by considering different possible $\mathrm{su(2)}$ representations embedded within the PXP model, we also identify a weak perturbation which enhances revivals from the $\vert \mathbb{Z}_3 \rangle = \vert 100100...\rangle$ state, and a strong deformation resulting in a new model which supports revivals from $\vert \mathbb{Z}_4 \rangle$ initial state. We also identify two deformations of the PXP model which fix an $\mathrm{su(2)}$ algebra \emph{completely}, such that the models feature exact wavefunction revivals from simple product states and an exact integrable Krylov subspace generated by repeated application of the Hamiltonian, while also simultaneously containing thermalizing sectors.

The remainder of this paper is organized as follows. Secs.~\ref{sec:pxp} and \ref{section:exact_embedding} contain an overview of the physics of the PXP model and the recent constructions of scarred models via projector embedding and dynamical symmetry. Sec.~\ref{sec:loose} introduces our notion of ``loose" embedding of broken Lie algebra representations into an eigenspectrum of a many-body system. In Sec.~\ref{sec:pxpz2} we present the simplest application of our construction to revivals from $|\mathbb{Z}_2\rangle$ product state in PXP model. %By imposing the commutation rules of $\mathrm{su(2)}$ Lie algebra, we derive the perturbations to PXP model that enhance the $\mathbb{Z}_2$ revivals. 
 In Sec.~\ref{sec:pxpz3} we explore a different $\mathrm{su(2)}$ Lie algebra representation which can be loosely embedded in the PXP model in order to give rise to revivals from $|\mathbb{Z}_3\rangle$ product state. Additionally, we find an exactly embedded subspace in a new model which represents a strong deformation of the PXP model.  In Sec.~\ref{sec:pxpz4} we demonstrate that our method can be used to stabilise revivals from $|\mathbb{Z}_4\rangle$ product state which are absent in the PXP model. Our conclusions are presented in Sec.~\ref{sec:conc}. Appendices contain a non-trivial perturbation that stabilizes $\mathbb{Z}_2$ revivals in the spin-1 generalization of the PXP model, as well as technical details on the second-order corrections to $\mathrm{su(2)}$ algebras.

\section{A brief overview of PXP model} \label{sec:pxp}

\begin{figure}
\includegraphics[scale=0.8]{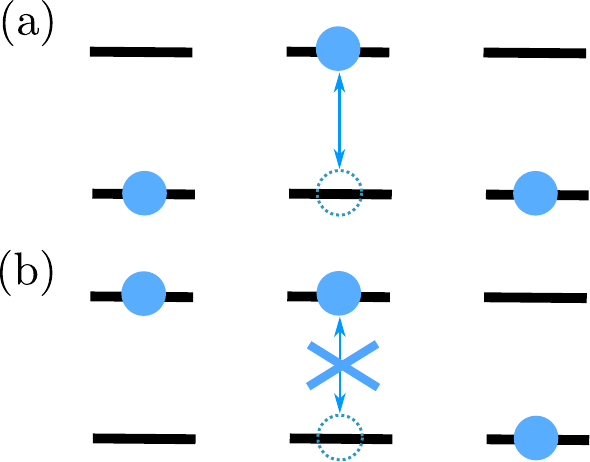}
\caption{An example of an allowed (a) and forbidden (b) transition under the Hamiltonian in Eq.~(\ref{eq:pxp}).}\label{fig:sketch}
\end{figure}

The PXP model~\cite{Lesanovsky2012} prevents adjacent excitations of atoms into the Rydberg states~\cite{Bernien2017}. The model can be expressed as a kinetically constrained spin-$1/2$ chain by denoting the basis of $\vert 0 \rangle = \vert {\downarrow} \rangle$, $\vert 1 \rangle = \vert {\uparrow} \rangle$, where $|0\rangle$ refers to an atom in its ground state and $|1\rangle$ denotes an excited state. The PXP Hamiltonian is given by
\begin{eqnarray}\label{eq:pxp}
    H_{\mathrm{PXP}} &=& \sum_{n=1}^N P_{n-1} \sigma^x_n P_{n+1},
\end{eqnarray}
where  $\sigma_n^x= \vert 0 \rangle_n \langle 1 \vert_n + \vert 1 \rangle_n \langle 0 \vert_n$ is the standard Pauli $x$-matrix on site $n$,  and the projector
\begin{equation}\label{eq:proj}
P_n = \vert 0 \rangle_n \langle 0 \vert_n
\end{equation}
implements correlated spin flips, i.e., $P$ removes any transitions that would create adjacent Rydberg excitations. Examples of allowed and forbidden processes are illustrated in Fig.~\ref{fig:sketch}.

Our numerical study of the model in Eq.~(\ref{eq:pxp}) and related models below will be based on exact diagonalization of finite chains with periodic boundary condition ($n+N \equiv n$).

The PXP model in Eq.~(\ref{eq:pxp}) is non-integrable and thermalizing~\cite{Turner2017}, but its quench dynamics is strongly sensitive to the choice of the initial state~\cite{Bernien2017}. For simplicity, we focus on initial states that are product states of atoms compatible with the Rydberg constraint (recent work in Ref.~\onlinecite{michailidis2017slow} studied the revivals from more general classes of weakly-entangled initial states). One such initial state is the N\'eel state $|\psi(0)\rangle = \vert \mathbb{Z}_2 \rangle \equiv \vert 0101... \rangle$, which gives rise to revivals in the quantum fidelity,
\begin{eqnarray}
\vert \langle \psi(0) \vert e^{-iHt} \vert \psi(0) \rangle \vert^2.
\end{eqnarray}
Other physical quantities, such as local observable expectation values, correlation functions as well as non-local quantities such as entanglement entropy, were all found to revive with the same frequency as the fidelity~\cite{TurnerPRB}. Other initial states such as $|\mathbb{Z}_3\rangle \equiv |100100\ldots\rangle$ also revive, though much more weakly, while states with larger unit cells, such as $|\mathbb{Z}_4\rangle \equiv |10001000\ldots\rangle$, do not revive even in small systems accessible by exact numerics~\cite{TurnerPRB}.

As we pointed out in the Introduction, the return probability of the 
$\vert \mathbb{Z}_2 \rangle$ state in PXP model  clearly decays with time, suggesting that the revival is fragile and likely to disappear in the thermodynamic limit. 
In this context, Ref.~\onlinecite{Choi2018} made an important observation that PXP model could be weakly deformed such that revivals are made nearly perfect. The enhancement of revivals in the PXP model  was explained by the fact that appropriate perturbations stabilise an approximate $\mathrm{su(2)}$ algebra formed by the special eigenstates of the PXP model. The special eigenstates can be described, with high accuracy, using a ``forward scattering approximation" (FSA)~\cite{Turner2017}. The FSA is based on a particular decomposition of the PXP Hamiltonian, $H_{\mathrm{PXP}} = H^+ + H^-$, chosen in such a way that $H^-$ annihilates the initial N\'eel state $|\mathbb{Z}_2 \rangle$ (with $H^+=(H^-)^\dagger$). The set of states $(H^+)^n |\mathbb{Z}_2 \rangle$ then form an orthogonal Krylov-like subspace of finite dimension $N+1$, where $N$ is the number of atoms. The scarred eigenstates can be compactly represented as linear superpositions of $N+1$ FSA basis states~\cite{TurnerPRB}. Within the subspace of special eigenstates, the operators $H^+$ and $H^-$ act like raising and lowering operators for a fictitious spin-$N/2$ particle. Intuitively, periodic revivals can then be interpreted as precession of this large spin~\cite{Choi2018}. In the pure PXP model, the emergent $\mathrm{su}(2)$  spin algebra is only approximate but becomes nearly exact at the optimal revival point.

 In this paper, we reinterpret the revivals in PXP model from the point of view of broken Lie algebras, by defining a set of broken generators for which the scar states act as an approximate basis. Considering corrections to this algebra allows us to construct perturbations that significantly enhance the revivals for general types of initial states without relying on FSA scheme.

\section{Exact Embedding of Scarred Eigenstates} \label{section:exact_embedding}

Before we consider PXP model which features approximate integrable subspaces with small subspace variance, which we term as having loosely embedded scar states, we first review several ways in which an \emph{exact} integrable subspace has been demonstrated to arise in recent works in the literature.

\subsection{Projector embedding} \label{section:ShiraishiScars}

Selected eigenstates can be embedded into the spectrum of an ergodic Hamiltonian via the ``projector embedding" construction due to Shiraishi and Mori~\cite{ShiraishiMori} (further extensions to topologically ordered systems have been developed in Ref.~\onlinecite{NeupertScars}). Consider a Hamiltonian describing some lattice system of the form:
\begin{equation}
H = \sum_{i=1}^N P_i h_i P_i + H^{\prime},
\end{equation}
where $P_i$ are arbitrary local projectors [not necessarily the same as in Eq.~(\ref{eq:proj})], $h_i$ are arbitrary local Hamiltonians acting on lattice sites $i=1,2,\ldots N$, $[H^{\prime},P_i] = 0$ for all  $i$, and $\vert \psi_i \rangle$ are target states that are annihilated by the projectors,
\begin{equation}
P_i \vert \psi_j \rangle = 0, \quad \forall \, i,j.
\end{equation} 
It follows 
\begin{eqnarray}
P_i H \vert \psi_j \rangle = P_i H^{\prime} \vert \psi_j \rangle = H^{\prime} P_i \vert \psi_j \rangle = 0,
\end{eqnarray}
thus $[H,P_i] = 0$ for all $i$, which implies $[H,\sum_i P_i]=0$. Therefore, $H$ takes the block diagonal form 
\begin{eqnarray}
H = H_{\mathrm{target}} \bigoplus H_{\bot},
\end{eqnarray}
where $H_{\mathrm{target}}$ is spanned by the target states $\vert \psi_i \rangle$. Such a decomposition may result in the model possessing both integrable and ergodic sectors. Models of this form generically contain eigenstates embedded near the center of the spectrum~\cite{ShiraishiMori}. There is no guarantee embedded states are equidistant in energy and may even be degenerate, such that this scheme can produce models which do not exhibit perfect wavefunction revivals. We note that, for periodic boundary conditions, the PXP model, introduced in Sec.~\ref{sec:pxpz2} below, can be expressed in this ``projector embedded" form such that a single target state is embedded -- namely the AKLT ground state at zero energy~\cite{Shiraishi_2019}. However, the complete set of $N+1$ scarred eigenstates with enhanced support on $\vert \mathbb{Z}_2 \rangle$ state (mentioned in Sec.~\ref{sec:pxp}) have not been understood through this embedding procedure.

\subsection{Equidistant embedding: Dynamical Symmetry} \label{section:IadecolaScars}

Next we review a way in which ETH violating eigenstates can be embedded with equidistant energy spacing, yielding exact wavefunction revivals in specially designed quenches. Consider a Hamiltonian of the form:
\begin{equation}
H = H_0 + H^{\prime}.
\end{equation}
We assume the existence of some local operator $Q^+$ for which there exists an extensive dynamical symmetry with $H^{\prime}$:
\begin{equation}
[H^{\prime}, Q^+] = \alpha Q^+,
\end{equation} 
such that, for any eigenstate $\vert \Omega \rangle$ of $H^{\prime}$, we can generate an equally spaced tower of eigenstates, $(Q^+)^n \vert \Omega \rangle$.
If the subspace given by a tower of $H^{\prime}$ eigenstates, $\vert n \rangle = 1/\mathcal{N} (Q^+)^n \vert \Omega \rangle$, are also zero energy eigenstates of $H_0$, $H^{\prime}$ will split the degeneracy such that $\vert n \rangle$ are equidistant eigenstates of the full Hamiltonian. Further, if $\vert \Omega \rangle$ is a weakly entangled state, due to the locality of $Q^{+}$, states $\vert n \rangle$ are also expected to be weakly entangled. Given an appropriate choice of $H_{0}$ such that the model is non-integrable, the states $\vert n \rangle$ will be weakly entangled scarred eigenstates which violate the ETH. Such a scenario has been realised in a variety of models, such as spin-1 XY models~\cite{Iadecola2019_2,Chattopadhyay}, a spin $1/2$ model with emergent kinetic constraints~\cite{Iadecola2019_3} and a spin chain where the dynamical symmetry emerges due to an underlying Onsager algebra~\cite{OnsagerScars}.

\begin{figure}
    \centering
    \includegraphics[width=\linewidth]{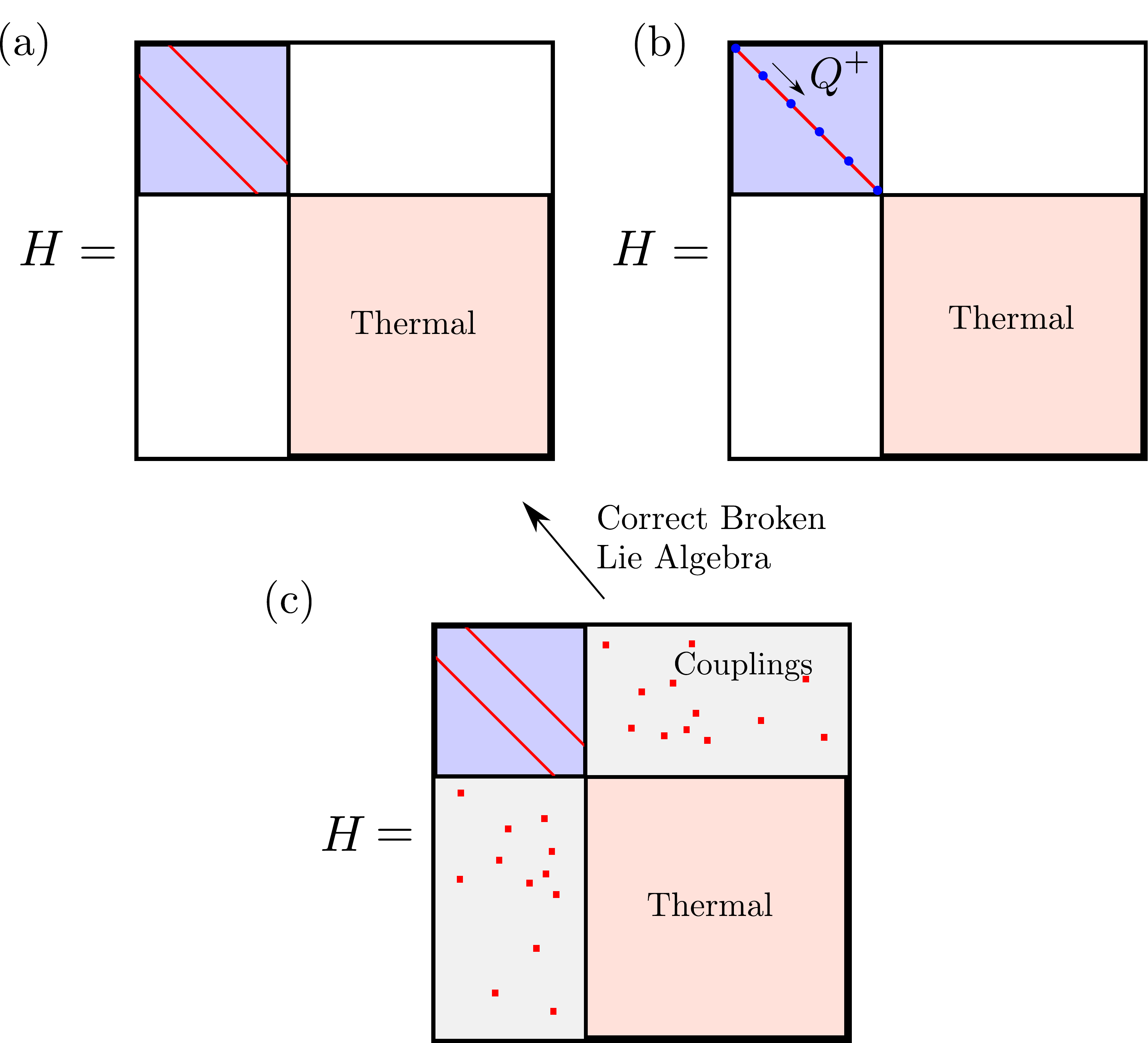}
    \caption{Summary of various mechanisms for embedding scarred eigenstates in a many-body system. 
   (a) An exactly embedded Krylov subspace (purple tridiagonal matrix, with red lines symbolizing the non-zero elements).    
   Such a scenario can emerge in models exhibiting the phenomenology of fractonic systems~\cite{Pretko2019}, where if the Krylov subspace is exponentially large this effect is coined ``Krylov-restricted thermalization"~\cite{MoudgalyaKrylov}.  Lifting the restriction the embedded subspace be tridiagonal, models of type (a) can also be generically realized by the ``projector embedding method" (Sec.~\ref{section:ShiraishiScars}).   
    (b) Exact scars featuring perfect revivals due to a dynamical symmetry of certain terms in the Hamiltonian generated by $Q^+$ (see Sec.~\ref{section:IadecolaScars}). 
        Type (b) scars have being realized in a variety of spin models such as spin-1 XY model~\cite{Iadecola2019_2,Iadecola2019_3,Chattopadhyay,OnsagerScars}. 
    (c) PXP-like scarring~\cite{Khemani2018,Choi2018}, where a Krylov subspace which approximately acts as an $\mathrm{su(2)}$ representation is sparsely coupled to the thermal bulk, such that the subspace has a low subspace variance (which is equivalent to the Frobenius norm of the block labelled couplings). By fixing various broken Lie algebra representations in models of type (c) we can also realize the scarred subspace of approximate type (a), where the nearly exactly embedded subspace forms a representation of the Lie algebra (as will be discussed in Sec.~\ref{section:pxpZ3Exact} and \ref{section:pxpZ4}).
    }
    \label{fig:scarMechs}
\end{figure}
A summary of exact embeddings is presented in Figs.~\ref{fig:scarMechs}(a), (b). In contrast to exact embeddings, the focus of this paper is the PXP model~\cite{Bernien2017} where the scarred subspace is only approximately decoupled from the thermal bulk, Fig.~\ref{fig:scarMechs}(c). Before discussing in detail the PXP model in Sec.~\ref{sec:pxpz2}, in the following section we introduce our general notion of loose embedding that can be applied, in principle, to any model.

\section{Loose embeddings of broken Lie algebras }\label{sec:loose}

Previous examples of exact embeddings of scarred eigenstates in Sec.~\ref{section:exact_embedding} are analytically tractable, but they do not directly apply to the experimentally observed scarred revivals in the PXP model~\cite{Bernien2017}. In the latter case, the revivals clearly decay over time, thus we are looking to interpret such revivals in terms of an \emph{inexact} embedding of an algebra whose representation is defined by the scarred states. Here we outline how to construct models with loosely embedded scar states, whose Hamiltonian approximately fractures into the form $H \approx H_{\mathrm{int}} \bigoplus H_{\bot}$, where $H_{\mathrm{int}}$ possesses an approximate dynamical symmetry, which we engineer from the root structure of a Lie algebra representations with weakly ``broken" commutation relations.

\subsection{Embedding scheme}\label{sec:scheme}

We start by recalling some basics of Lie algebras and representation theory. 
Infinitesimal generators $g_i$ of a Lie group $\mathcal{G}$ form  a Lie Algebra $\mathcal{A}$:
\begin{eqnarray}
[g_i,g_j] = f^k_{ij} g_k.
\end{eqnarray}
The algebra is encoded in the structure constants $f^k_{ij}$, which are antisymmetric with respect to lower indices, $f^k_{ij} = - f^k_{ji}$. 
A set of $n \times n$ matrices $\{M_i\}$ satisfying $[M_i,M_j] = f^k_{ij} M_k$ forms an $n$-dimensional representation of the Lie algebra. Verifying these commutation relations is sufficient to verify the set $\{M_i\}$ forms a valid representation.

%\subsection{Cartan-Weyl Basis, Roots of Lie Algebra}
Given a set of infinitesimal generators of a Lie group, define $\{H^i\}$ as the largest set of mutually commuting generators. By taking linear combinations of the remaining generators, one can construct a set of ladder operators, $\{E^\alpha\}$:
\begin{eqnarray}
     [H^i, E^\alpha] = \alpha^i E^\alpha. 
     \label{eq:root_system}
\end{eqnarray}
Together, the sets $\{H^i\}, \{E^{\alpha}\}$ are known as the Cartan-Weyl basis.  As the set $\{H^i\}$ are mutually commuting by definition, there exists a basis which simultaneously diagonalizes every $H^i$ such that we can label basis states of a representation by their $H^i$ quantum numbers. On application of $E^{\alpha}$, the change in $H^i$ quantum numbers is just the roots $\alpha^i$:
\begin{eqnarray}
    H^i \vert \psi \rangle &=& \lambda_i \vert \psi \rangle, \\
    H^i E^{\alpha} \vert \psi \rangle &=& (E^\alpha H^i + \alpha^i E^\alpha) \vert \psi \rangle =  (\lambda_i + \alpha^i) E^\alpha \vert \psi \rangle. \quad
\end{eqnarray}
Given a single basis state which is an eigenstate of every $H^i$, one can systematically construct the remaining basis states via repeated applications of the ladder operators $E^\alpha$. This construction will prove useful for forming approximate basis states of broken Lie algebra representations, which can be used to approximate many-body scar states (e.g., within the FSA scheme~\cite{Turner2017}).

Consider the set of operators $\{E^\alpha\}$ which are raising and lowering operators of some Lie algebra $\mathcal{A}$ in the Cartan-Weyl basis. The set of equations,
\begin{eqnarray}
 [E^\alpha,E^\beta] = \sum_\gamma c_{\gamma}E^\gamma + \sum_i d_i H^i,
 \label{eq:Hi_def}
\end{eqnarray}
follows from the properties of the Lie algebra, but can be taken as defining the operators $H^i$ when these equations are inverted.

Now we are in position to introduce our notion of ``broken" Lie algebra. Let the set of operators $\{\bar{E}^\alpha\}$ be of equal size as the previous set $\{ E^\alpha \}$, but we do not assume they are raising/lowering operators of any Lie algebra. Taking Eqs.~(\ref{eq:Hi_def}) as a definition for $H^i$ as some linear combination of $\{E^\alpha, [E^\alpha,E^\beta]\}$, define $\bar{H^i}$ as the same linear combination of $\{\bar{E}^\alpha,[\bar{E}^\alpha,\bar{E}^\beta]\}$.

If the sets $\{\bar{E}^\alpha\}$, $\{\bar{H}^i\}$ satisfy:
\begin{eqnarray}
 [\bar{H}^i, \bar{E}^\alpha] = \alpha^i \bar{E}^\alpha + \delta^\alpha, 
\end{eqnarray}
where $\alpha^i$ are the root coefficients of the Lie algebra $\mathcal{A}$ and it is understood $\delta^{\alpha}$ contain no terms proportional to the generators $\bar{E}^{\alpha}$, we say $\{\bar{E}^\alpha\}, \{\bar{H}^i\}$ form a \emph{broken representation} of the Lie algebra $\mathcal{A}$. 

Now consider a Hamiltonian consisting of a linear combination of the diagonal generators $\{\bar{H}^i\}$ rotated to some other basis:
\begin{eqnarray}
    H = \sum_n a_n U^{\dagger} \bar{H}^n U,
      %= \sum_n \beta_n \bar{H}_n + \sum_m \gamma_m \bar{E}^{m},
    \label{eq:Hamiltonian_lin_comb}
\end{eqnarray}
where $U$ is an arbitrary unitary rotation.  Consider quenching from a
simultaneous eigenstate $\vert \psi_0 \rangle$ of the operators
$\{\bar{H}^i\}$. Construct an approximate basis for the broken
representation by repeated application of the raising operators
$\bar{E}^\alpha$ on $\vert \psi_0 \rangle$. If the algebra were exact, the
Hamiltonian would fracture into the block diagonal form $H = H_{\mathrm{rep \,\,
basis}} \bigoplus H_{\bot}$ and there would exist several dynamical symmetries of $H_{\mathrm{rep \,\, basis}}$, corresponding to the rotated ladder operators, $Q_{\alpha} = U^{\dagger} E^{\alpha} U$. 
For a broken Lie algebra, these relations become approximate, thus Hamiltonians of the form of Eq.~(\ref{eq:Hamiltonian_lin_comb}) will contain an approximate dynamical symmetry within a loosely embedded integrable subspace.

It is possible the dynamics can resemble a quench with additional decoherence from the related system $H(\bar{H^i},\bar{E}^{\alpha}) \rightarrow H(H^i,E^{\alpha} $). For example, if the embedded algebra was $\mathrm{su(2)}$, it is possible the wavefunction will revive with a single frequency provided the following conditions are met:
\begin{enumerate}
    \item The variance of the approximate basis with respect to $\bar{H}^i$ is sufficiently small.
    \item The spacing of expectation values with respect to $\bar{H}^i$ after applications of $\bar{E^\alpha}$ to $\vert \psi_0 \rangle$ approximately obeys the root structure of the desired Lie algebra, i.e., 
    \begin{eqnarray}
    \frac{\langle \phi \vert \bar{H}^i \vert \phi \rangle}{\langle \phi \vert \phi \rangle} &\approx \lambda_i + \alpha^i,
    \end{eqnarray}
where $ \bar{H}^i \vert \psi_0 \rangle = \lambda_i \vert \psi_0 \rangle$ and 
         $   \vert \phi \rangle = \bar{E}^\alpha \vert \psi_0 \rangle$.
    \item Repeated application of $\bar{E}^\alpha$ on $\vert \psi_0 \rangle$ will terminate after a finite number of steps, thus generating a subspace of the full Hilbert space. In general, this subspace does not correspond to an exact symmetry sector of the Hamiltonian. To see signatures of the exact Lie algebra, this subspace must be sufficiently disconnected from the orthogonal space under the action of the Hamiltonian.
\end{enumerate}

\subsection{Iterative corrections to broken Lie algebras: Identifying perturbations that stabilize revivals}

By perturbing the operators $\bar{E^\alpha}$ with terms that appear in the error $\delta^\alpha$, it is possible to improve the broken Lie algebra, in the sense that decoherence in the previously described quench in Sec.~\ref{sec:scheme} is reduced. 

\begin{figure}[htb]
    \centering
    \includegraphics[width=0.5\textwidth]{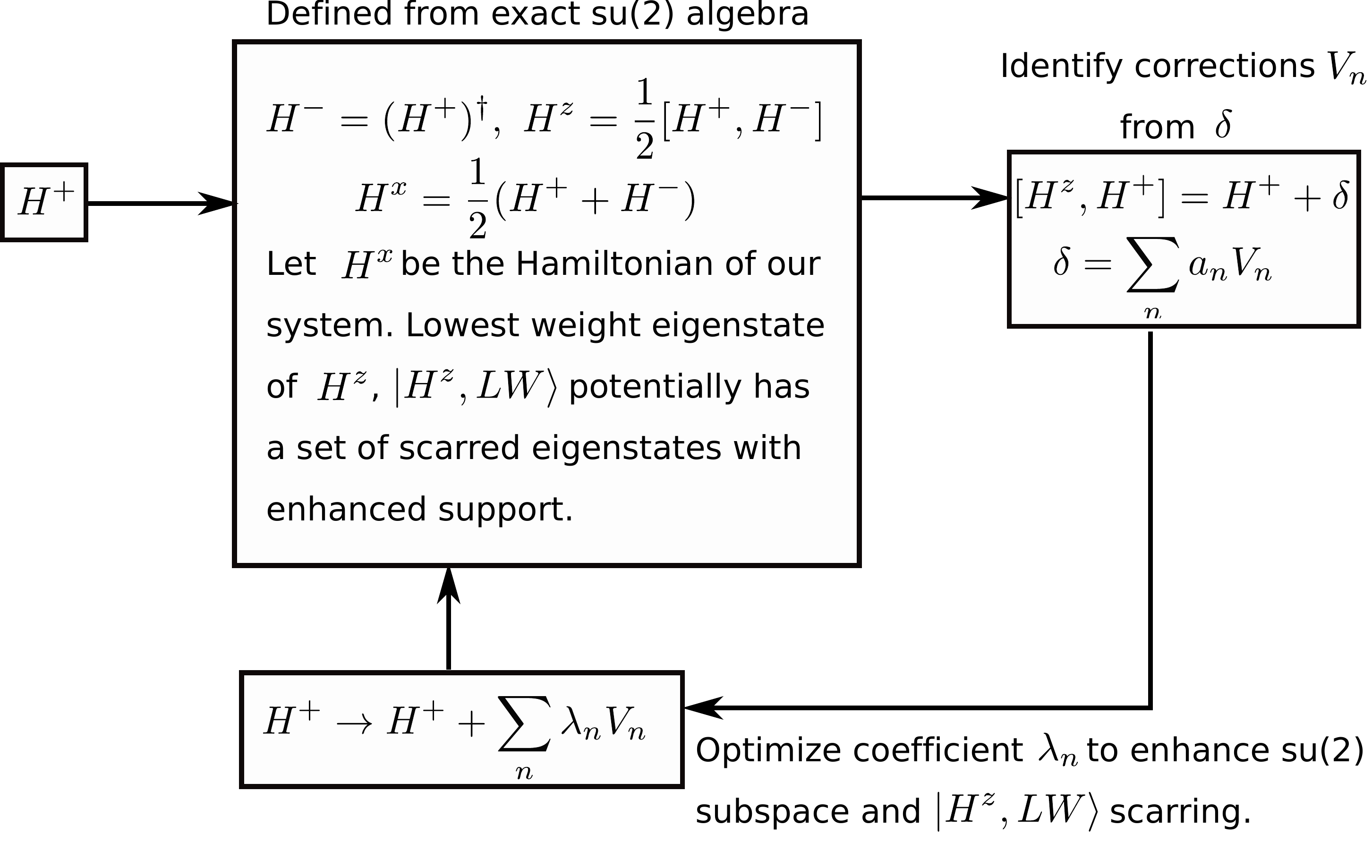}
    \caption{Schematic illustration of our iterative scheme which identifies corrections to broken Lie algebras, specifically an $\mathrm{su(2)}$ Lie algebra in this case. The optimization of $\lambda_n$ is with respect to the error measures described in the text, such as maximizing the first fidelity peak $|\langle H^z, LW \vert e^{-iHt} \vert H^z, LW \rangle|^2$ or minimizing the subspace variance of $H$ w.r.t. to the $\mathrm{su(2)}$ basis defined in Eq.~(\ref{eq:lw}). %$\vert n \rangle = 1/\mathcal{N} (H^+)^n \vert H^z, LW \rangle$. 
    }
    \label{fig:algebraCorrection}
\end{figure}

Consider some broken representation of a Lie algebra:
\begin{eqnarray}
    [\bar{H}^i, \bar{E}^\alpha] = \alpha^i \bar{E}^{\alpha} + \delta^\alpha, \quad
   \delta^\alpha = \sum_n a_n V_n^\alpha,
\end{eqnarray}
where the error $\delta^\alpha$ has been decomposed into terms sharing the same coefficient $a_n$. Now perturb the raising/lowering operators as follows:
\begin{eqnarray}
\bar{E}^\alpha_{(1)} = \bar{E}^\alpha + \sum_n c_n V_n^\alpha. 
\end{eqnarray}
This in turns defines new $\bar{H}^i_{(1)} = \bar{H}^i + H^i_{\mathrm{perts}}$ , following the same definition of $H^i$ in Eq.~(\ref{eq:Hi_def}). It follows:
\begin{align}
    [ {\bar{H}^i}_{(1)},{\bar{E}^\alpha}_{(1)}] &= \alpha^i \bar{E}^\alpha + \sum_m f_m(c_0,...,c_N) V_m^\alpha + {\delta^\alpha}_{(2)},  \\
    {\delta^\alpha}_{(2)} &= \sum_n g_n(c_0,...,c_N) {V^\alpha}_{(2)n},
    \label{eq:polynomial}
 \end{align}
 where $f_m(c_0,...,c_N)$, $g_n(c_0,...,c_N)$ are polynomials in the perturbation coefficients and ${V^\alpha}_{(2)n}$ are second order error terms. If the coefficients $c_n$ can be optimized to satisfy
 \begin{eqnarray}
 [\bar{H^i}_{(1)}, \bar{E}^\alpha_{(1)}] \approx \alpha^i \bar{E}^\alpha_{(1)} + \delta^\alpha_{(2)},
 \end{eqnarray}
 such that decoherence in the previously described quench is reduced, we say that the broken representation has been improved. This can lead to decreased variance of $\{H^i\}$ and/or improved spacing of $\langle H^i \rangle$ with respect to the approximate basis of the broken representation and also may result in the approximate basis becoming more disconnected from the orthogonal subspace under the action of the perturbed Hamiltonian [Eq.~\ref{eq:Hamiltonian_lin_comb}, with $H(H^i, E^{\alpha}) \rightarrow H(H^i_{(1)}, E^{\alpha}_{(1)})$]. Further, if the representation improves, we expect the magnitude of the error terms to decrease, given by the Frobenius norm $\vert \vert \delta^\alpha_{(2)} \vert \vert_F < \vert \vert \delta^\alpha \vert \vert_F$. Fig.~\ref{fig:algebraCorrection} schematically shows this process of identifying corrections to the algebra. We will demonstrate that this procedure results in many-body scarred models with long-lived coherent dynamics in the subsequent sections.
 
Before illustrating this approach with examples of a broken $\mathrm{su(2)}$ Lie algebra, we briefly discuss ways of quantifying how much the approximate  $\mathrm{su(2)}$ Lie algebra representation differs from an exact representation.  As a possible error measure, we consider $\mathrm{max} \; \mathrm{var}(\bar{H}^z)_n$ with respect to the approximate basis, where $\mathrm{var}(\bar{H}^z)_n$ is the variance of the basis state $\vert n \rangle$ defined as
\begin{eqnarray}\label{eq:lw}
\vert n \rangle = \frac{1}{\sqrt{\mathcal{N}}} (\bar{H}^+)^n \vert \mathrm{LW} \rangle,
\end{eqnarray}

with $| \mathrm{LW}\rangle$ being the lowest weight state of the $\mathrm{su(2)}$ Lie algebra representation $\vert S,-S \rangle$. This state obeys $\bar{H}^-\ket{\mathrm{LW}}{=}0$, or equivalently, it is the ground state of $\bar{H}^z$.
If the revivals are due to an $\mathrm{su}(2)$ algebra, we expect the corresponding basis states should have harmonic (equal) energy spacing. To quantify the deviation from harmonic spacing we introduce the quantity $K$:
\begin{eqnarray}
K = \vert \vert M \vert \vert_F, \quad  M_{nm} = \vert \Delta E_n - \Delta E_m \vert,
\end{eqnarray}
which represents the Frobenius norm of the matrix of level spacings. The latter are given by 
\begin{eqnarray}
\Delta E_n &= \langle \bar{H}^z \rangle_{n+1} - \langle \bar{H}^z \rangle_n, \quad \langle \bar{H}^z \rangle_n &= \langle n \vert \bar{H}^z \vert n \rangle.
\end{eqnarray}
To quantify how disconnected the subspace spanned by $\vert n \rangle$ is from  its orthogonal subspace under the action of the Hamiltonian, we use the subspace variance $\sigma$:
        \begin{eqnarray}
            \sigma = \mathrm{tr}\Big( (U_{\mathrm{rep}}^{\dagger} H^2 U_{\mathrm{rep}}) - ( U_{\mathrm{rep}}^{\dagger} H U_{\mathrm{rep}} )^2 \Big),         
        \end{eqnarray}
        where $U_{\mathrm{rep}}$ is the unitary operator which projects to the broken representation basis. This quantity can be interpreted as being proportional to the Frobenius norm of the block labelled couplings in Fig~\ref{fig:scarMechs}(c).

\section{Example: PXP model and embedded $\mathrm{su(2)}$ algebra }\label{sec:pxpz2}

We now exemplify our general embedding scheme outlined in Sec.~\ref{sec:loose} by using the PXP model~\cite{Lesanovsky2012,Bernien2017}. We demonstrate how to identify and improve the broken $\mathrm{su(2)}$ algebra associated with $\mathbb{Z}_2$ revivals.

\subsection{$\mathbb{Z}_2$ revivals and $\mathrm{su}(2)$ algebra}

First we focus on the well-known case of $\mathbb{Z}_2$ revivals in the PXP model~\cite{Choi2018}. Define the $\mathrm{su}(2)$ spin raising operator 
\begin{eqnarray}
    \bar{H}^+  \equiv \sum_n \left(  \tilde{\sigma}^+_{2n} + \tilde{\sigma}^-_{2n-1} \right),
\end{eqnarray}
where we have introduced the shorthand notation
\begin{eqnarray}\label{eq:sigmatilde}
    \tilde{\sigma}^{\alpha}_n \equiv P_{n-1} \sigma^\alpha_n P_{n+1}.
\end{eqnarray}
We have $H_{\mathrm{PXP}} = \bar{H}^+ + \bar{H}^-$ such that $H_{\mathrm{PXP}} = \bar{H}^x$ can be interpreted as an element of  su(2) algebra.
From the commutation rules of $\mathrm{su}(2)$ algebra, the diagonal element is given by (half) the commutator (note the minus sign)
\begin{eqnarray}
    \bar{H}^z  \equiv \frac{1}{2} [\bar{H}^+,\bar{H}^-] = \sum_n \left( \tilde{\sigma}^z_{2n} - \tilde{\sigma}^z_{2n-1} \right).
\end{eqnarray}
The reason for this choice of $\bar{H}^{+/-}$ is that the lowest weight state of $\bar{H}^z$ is the N\'eel state, $\vert 0101...\rangle$. 
We seek a representation for which $|\mathbb{Z}_2\rangle$ is the lowest weight state of $\bar{H}^z$ as, for an exact algebra, the lowest/highest weight states of $\bar{H}^z$ are also simultaneously eigenstates of the Casimir operator, such that repeated application of $\bar{H}^+$ on the lowest weight state would generate an $\mathrm{su(2)}$ subspace. To be explicit, consider the exact algebra $H^+ = \sum_n \sigma_n^+$, $H^- = (H^+)^{\dagger}$, $H^z = \sum_n \sigma_n^z$. Of the eigenstates of $H^z$, only repeated application of $H^+$ on the lowest weight state $\vert 000...\rangle = \vert S=N/2,S_z=-N/2 \rangle$ would generate an $\mathrm{su(2)}$ subspace. Superpositions of states with equal number of singlets must be taken as the root state for which repeated application of $H^+$ would generate further $\mathrm{su(2)}$ sectors.
 
It further follows:
\begin{eqnarray}
    \left[ \bar{H}^z,\bar{H}^{+} \right] &=& \bar{H}^{+} + \delta^{+}_{(1)}, \label{eq:delta1p} \\
    \left[ \bar{H}^z, \bar{H}^{-} \right]&=& -\bar{H}^{-} + \delta^{-}_{(1)}, \label{eq:delta1m}
\end{eqnarray}
where the error terms that break the algebra are
\begin{eqnarray}
\nonumber \delta^+_{(1)} &=& -\frac{1}{2} (PP\sigma^+_{2n}P + P\sigma^+_{2n}PP \\
                         &+&  P \sigma^-_{2n+1}PP + PP \sigma^-_{2n+1}P), \label{eq:z2FirstOrder_raising} \\
 \nonumber \delta^-_{(1)} &=&  \frac{1}{2} (PP\sigma^-_{2n}P + P\sigma^-_{2n}PP \\
  &+& P \sigma^+_{2n+1}PP + PP \sigma^+_{2n+1}P).
\end{eqnarray}
For brevity, we have suppressed a summation over the lattice sites in the definition of $\delta_{(1)}^{+/-}$, and terms like $PP\sigma_{2n}^+P$ stand for $\sum_n P_{2n-2}P_{2n-1}\sigma_{2n}^+P_{2n+1}$ (i.e., strings of $P$'s act on consecutive neighboring sites). 

\begin{figure}
    \centering
    \includegraphics[width=0.5\textwidth]{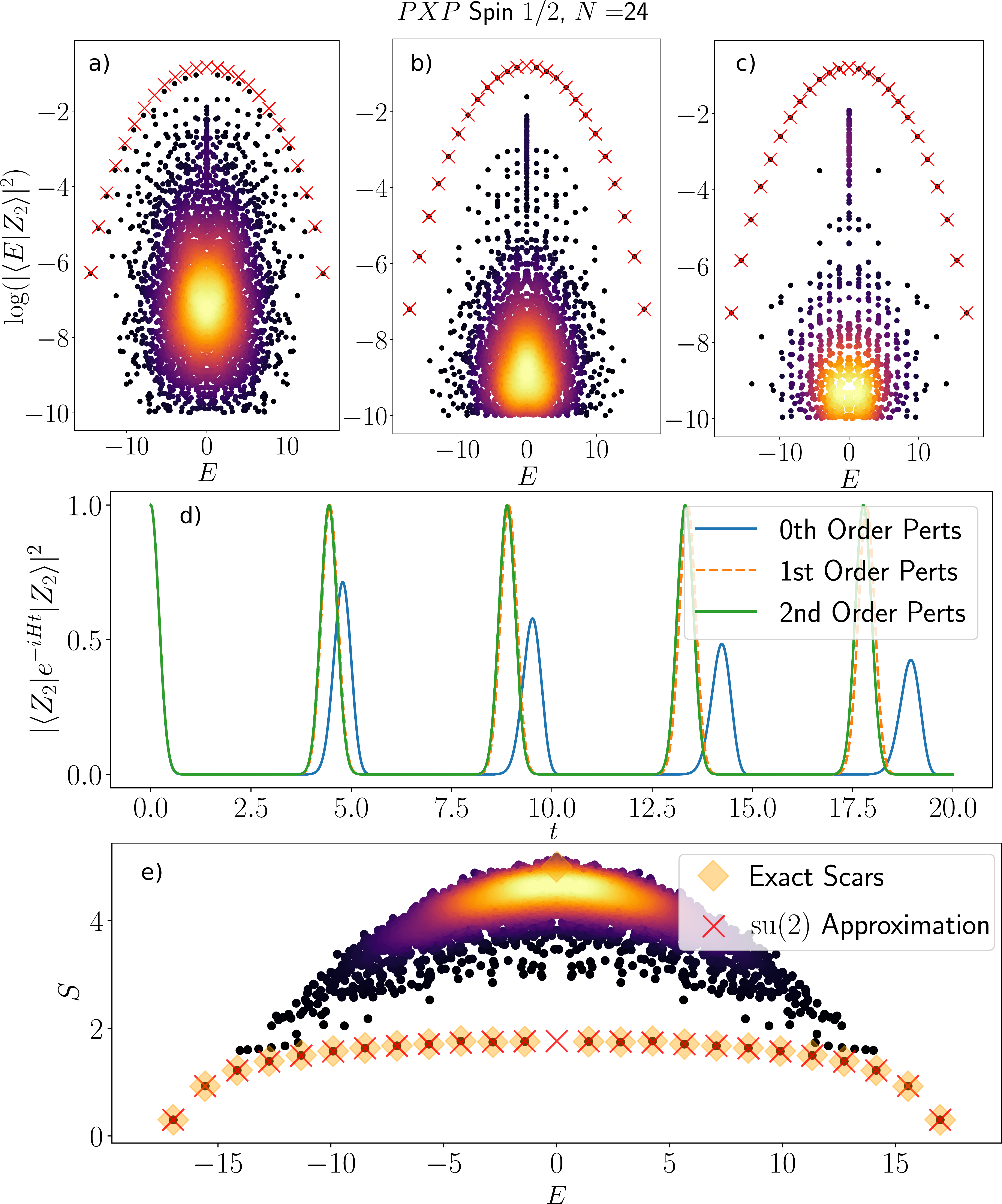}
    \caption{$\mathbb{Z}_2$ revival in PXP model. (a) Eigenstate overlap with the N\'eel $|\mathbb{Z}_2\rangle$ state. (b) Eigenstate overlap after including the first order $\mathrm{su}(2)$ correction (Eq.~\ref{eq:pxppert1}). (c) Eigenstate overlap after including the second order $\mathrm{su}(2)$ correction (Eq.~\ref{eq:z2_2ndOrder_1_1}-\ref{eq:z2_2ndOrder_1_N}). (d) Quantum fidelity in $\mathbb{Z}_2$ quench, with and without perturbations. Perturbation coefficients are those that maximize the first fidelity revival peak. (e) Bipartite entropy, Eq.~(\ref{eq:entropy}), of the eigenstates of PXP model after including second order $\mathbb{Z}_2$ $\mathrm{su(2)}$ corrections. The states labelled ``Exact Scars" are exact diagonalization results identified from the top band of states in (c). Red crosses in (a), (b), (c), (e) indicate approximate scar states obtained by projecting the Hamiltonian to the broken $\mathrm{su}(2)$ basis and diagonalizing.  Color scale in (a), (b), (c), (e) indicates the density of data points, with lighter regions being more dense.}
    \label{fig:z2pxpHalfSumm}
\end{figure}

From the expressions in Eqs.~(\ref{eq:delta1p})-(\ref{eq:delta1m}), we see that $\{ \bar{H}^z,\bar{H}^+,\bar{H}^- \}$ form a broken representation of $\mathrm{su}(2)$. In this language, the forward scattering approximation (FSA)~\cite{Turner2017} is rephrased as projecting the Hamiltonian $H$ to the broken representation basis in Eq.~(\ref{eq:lw}), with $|\mathrm{LW}\rangle \equiv |\mathbb{Z}_2\rangle$,
and diagonalizing. This procedure gives very accurate approximations to the special eigenstates of the full PXP model -- see red crosses in Fig.~\ref{fig:z2pxpHalfSumm} (a), (b), (c), (e).
\begin{table}[t]
    \begin{tabular}{|c|c|c|c|c|}
        \hline
        Order & $1-f_0$ & $\sigma/D_{\mathrm{su(2)}}$ & $max(var(H^z)_n)$ & $K$ \\
        \hline
        $n=0$ & $2.853 {\times} 10^{-1}$ & $1.116 {\times} 10^{-1}$ & $2.711 {\times} 10^{-1}$ & $9.310 {\times} 10^{0}$ \\
        \hline
        $n=1$ & $6.760 {\times} 10^{-4}$ & $2.190 {\times} 10^{-4}$ & $9.694 {\times} 10^{-4}$ & $6.008 {\times} 10^{-1}$ \\
        \hline
        $n=2$ & $3.113 {\times} 10^{-6}$ & $3.303 {\times} 10^{-6}$ & $2.355 {\times} 10^{-5}$ & $8.090 {\times} 10^{-2}$ \\
        \hline
    \end{tabular}
    \caption{Error metrics for the $\mathbb{Z}_2$ $\mathrm{su(2)}$ subspace of the PXP model at various perturbation orders for $N=24$. Subspace variance $\sigma$ is normalized by the dimension of the $\mathrm{su(2)}$ representation, $N+1$. See text for details of the pertubations.}
    \label{tab:pxpz2}
\end{table}

Next, we continue our program and identify a perturbation which can potentially improve the $\mathrm{su}(2)$ representation. First, define $\bar{H}^\pm_{(1)} = \bar{H}^\pm + \lambda \delta^\pm_{(1)}$. This gives us

\begin{eqnarray} 
\nonumber         H_{(1)} &=& H + \lambda(\delta^+_{(1)} + \delta^-_{(1)})  \\
         &=&  P \sigma_n^x P  + \lambda (P \sigma_n^x PP + PP \sigma_n^x P).
         \label{eq:pxppert1}
\end{eqnarray}
In order to find the optimal perturbation strength $\lambda$, we maximize the first fidelity revival as a function of $\lambda$,
\begin{eqnarray}
f_0(\lambda) = f(\lambda, t_0) = \vert \langle \psi(0) \vert e^{-iH(\lambda)t_0} \vert \psi(0) \rangle \vert^2,
\end{eqnarray}
where $t_0$ is the time at which the first revival occurs. Note that $t_0$ is $\lambda$-dependent. Throughout this paper, the minimization was carried out using the Python SciPy routine that employs the ``Sequential Least Squares Programming" (SLSQP) method. After optimization, we recover the perturbation that was previously empirically found~\cite{Khemani2018} to enhance the revivals following a $\vert \mathbb{Z}_2 \rangle$ quench with maximal $f_0$ when $\lambda = 0.108$ (at system size $N=18$). It was previously demonstrated the PXP model remains non-integrable after including this perturbation~\cite{Choi2018}. Note that the first order perturbation improves all error metrics of the broken representation, see Table.~\ref{tab:pxpz2}. 

Second order perturbations can be obtained in a similar fashion, although algebraic manipulations become very laborious to perform by hand. Our analytical results have been tested against a custom-designed software for symbolic computations of the nested commutators involving projectors~\footnote{K. Bull (\href{https://github.com/Cable273/comP}{https://github.com/Cable273/comP}).}. Fig.~\ref{fig:z2pxpHalfSumm} summarizes the differences between models after including first and second order perturbations. We find the scarred eigenstates become increasingly decoupled from the thermal bulk and can also be characterized by their anomalously low bipartite entanglement entropy $S$, defined in the usual way
\begin{eqnarray}
S &=& -\Tr(\rho_A \ln \rho_A), \label{eq:entropy}
\end{eqnarray}
in terms of the reduced density matrix $\rho_A = \Tr_{B} |\psi\rangle \langle \psi|$,  obtained via partial trace over the subsystem $B$ for some bipartition of the total system into two halves, $A$ and $B$, in the computational basis.

Restricting to terms with only a single spin flip, we identify the following second order error terms $\delta^+_{(2)}$:
\begin{eqnarray}
\nonumber \delta^+_{(2),1} &=&  P\sigma^z P \sigma_{2n}^+P + P \sigma_{2n}^+ P \sigma^z P 
\label{eq:z2_2ndOrder_1_1} \\
&+&  P \sigma^z P \sigma_{2n+1}^- P + P \sigma_{2n+1}^- P \sigma^z P, \\
\nonumber \delta^+_{(2),2} &=& P \sigma_{2n}^+ PPP + PPP \sigma_{2n}^+ P \\
&+&  P \sigma_{2n+1}^- PPP + PPP \sigma_{2n+1}^- P, \\
 \delta^+_{(2),3} &=& PP \sigma_{2n}^+ PP + PP \sigma_{2n+1}^- PP, \\
\nonumber \delta^+_{(2),4} &=&  PP \sigma_{2n}^+ P \sigma^z P + P \sigma^z P \sigma_{2n}^+ PP \\
&+& PP \sigma_{2n+1}^- P \sigma^z P + P \sigma^z P \sigma_{2n+1}^- PP, \\
\nonumber \delta^+_{(2),5} &=& PPP \sigma_{2n}^+ PP + PP \sigma_{2n}^+ PPP \\ 
&+& PPP \sigma_{2n+1}^- PP + PP \sigma_{2n+1}^- PPP, \\
\nonumber \delta^+_{(2),6} &=& P \sigma_{2n}^+ P \sigma^z PP + PP \sigma^z P \sigma_{2n}^+ P \\
&+& PP \sigma^z P \sigma_{2n+1}^- P + P \sigma_{2n+1}^- P \sigma^z PP, \\
\nonumber \delta^+_{(2),7} &=& PPPP \sigma_{2n}^+ P + P \sigma_{2n}^+ PPPP \\
&+& PPPP \sigma_{2n+1}^- P +  P \sigma_{2n+1}^- PPPP , \\
\nonumber \delta^+_{(2),8} &=& PP \sigma_{2n}^+ P \sigma^z PP + PP \sigma^z P \sigma_{2n}^+ PP \\
&+& PP \sigma_{2n+1}^- P \sigma^z PP + PP \sigma^z P \sigma_{2n+1}^- PP. \label{eq:z2_2ndOrder_1_N}
\label{eq:z2_2ndOrder_2}
\end{eqnarray}
Putting these terms together, we obtain the second order perturbations, 
$\bar{H}^+_{(2)} = \bar{H}^+ + \lambda_0 \delta^+_{(1)} + \sum_{i=1}^8 \lambda_i \delta^+_{(2),i}$ and $\bar{H}^-_{(2)} = \bar{H}^- + \lambda_0 \delta^-_{(1)} + \sum_{i=1}^8 \lambda_i \delta^-_{(2),i}$,  which in turn define $H_{(2)} = \bar{H}^+_{(2)} + \bar{H}^-_{(2)}$. Coefficients optimizing fidelity were found to be:
\begin{eqnarray}
    \lambda_i^* &= [0.11135,0.000217,-0.000287,-0.00717,\\
&0.00827,0.00336,0.00429,0.0103,0.00118],
\end{eqnarray}
where the first value is the optimal coefficient for the first order term Eq.~(\ref{eq:z2FirstOrder_raising}), while the remaining coefficients correspond to the terms in order of appearance in Eqs.~(\ref{eq:z2_2ndOrder_1_1})-(\ref{eq:z2_2ndOrder_1_N}).
These values have been found via numerical optimization at system size $N=16$. Note that previous work in Ref.~\cite{Choi2018} only considered $PXPIP+PIPXP$ as a second order perturbation to $H_{\mathrm{PXP}}$. By including all spin flip terms obtained from the Lie algebra error, fidelity can be enhanced to $1-f_0 \approx O(10^{-6})$, while  if we only retain  $PXPIP+PIPXP$ we obtain infidelity that is a few orders of magnitude higher, $1-f_0 \approx O(10^{-3})$  (data for $N=16$). In Ref.~\onlinecite{Choi2018} fidelity on the order $1-f_0 \approx O(10^{-6})$ was found by including only terms $P_{n-1}X_nP_{n+1}P_{n+d}+P_{n-d}P_{n-1}X_nP_{n+1}$ up to high order $d \leq 10$, which are expected to arise as corrections in higher orders of our method. While these terms alone appear sufficient to reach very high fidelity values, our analysis suggests that, strictly speaking, these terms do not fully fix the $\mathrm{su(2)}$ algebra. 

The decomposition of $H_{\mathrm{PXP}}=\bar{H}^++\bar{H}^-$ used to identify the broken $\mathrm{su(2)}$ algebra assosciated with $\mathbb{Z}_2$ revivals is not unique. In the following Sections, we discuss further decompositions leading to additional $\mathrm{su(2)}$ representations which can be enhanced to fix revivals from $\vert \mathbb{Z}_3 \rangle$ and $\vert \mathbb{Z}_4 \rangle$ initial states.

\section{$\mathbb{Z}_3$ revivals from $\mathrm{su}(2)$ algebra}
\label{sec:pxpz3}

In addition to $\mathbb{Z}_2$ revivals, PXP model was also shown numerically to exhibit wave function revivals following a quench from $\vert \mathbb{Z}_3 \rangle = \vert 100100...\rangle$ state~\cite{Turner2017, TurnerPRB}. (Somewhat more robust revivals are in fact seen from a weakly-entangled initial state ``close" to $|\mathbb{Z}_3\rangle$~\cite{Michailidis2019}.) Unlike $\mathbb{Z}_2$ state, the revivals from $\mathbb{Z}_3$ sharply decay even in numerical simulations on fairly small systems~\cite{TurnerPRB}, suggesting the model is even further away from any exact Lie algebra representation furnished by $|\mathbb{Z}_3\rangle$ state. 

The $\mathbb{Z}_3$ revivals originate from $2N/3+1$ scarred eigenstates with enhanced support on the $\vert \mathbb{Z}_3 \rangle$ state. We stress that out of these $2N/3+1$ scarred eigenstates, only two eigenstates coincide with the $N+1$ scarred eigenstates with enhanced support on $\mathbb{Z}_2$, which are the ground and most excited eigenstates of the model. Thus, we interpret the $\mathbb{Z}_3$ scarred subspace as a distinct loosely embedded $\mathrm{su(2)}$ subspace as compared to the $\mathbb{Z}_2$ scarred subspace. There has been no FSA method to describe the $2N/3+1$ $\mathbb{Z}_3$ scar states and, consequently, the perturbations that improve the $\mathbb{Z}_3$ revival are not known. Here we demonstrate that it is possible to deform the PXP model to stabilise a \emph{different}  $\mathrm{su}(2)$ algebra representation compared to the $\mathbb{Z}_2$ case, which results in robust $\mathbb{Z}_3$ revivals. 

We follow our general approach and start by introducing raising and lowering operators compatible with $|\mathbb{Z}_3\rangle$ state:
\begin{eqnarray}
    \bar{H}^+ &=& \sum_n \left( \tilde{\sigma}^-_{3n} + \tilde{\sigma}^+_{3n+1} + \tilde{\sigma}^+_{3n+2} \right), \label{eq:z3raising} \\
    \bar{H}^- &=& \sum_n \left( \tilde{\sigma}^+_{3n} + \tilde{\sigma}^-_{3n+1} + \tilde{\sigma}^-_{3n+2}\right),
\end{eqnarray}
where, as before, we have $H_{\mathrm{PXP}} = \bar{H}^+ + \bar{H}^-$. The $\mathrm{su}(2)$ diagonal generator is then given by $\bar{H}^z = \frac{1}{2} [\bar{H}^+,\bar{H}^-]$,
which can be shown to take the form
\begin{eqnarray}
    \nonumber \bar{H}^z &=& \sum_n -\tilde{\sigma}^z_{3n} + \tilde{\sigma}^z_{3n+1} + \tilde{\sigma}^z_{3n+2} \\
\nonumber &+& \frac{1}{2}\sum_n \Big(  P_{3n} \sigma_{3n+1}^+ \sigma_{3n+2}^- P_{3n+3} \\
&+& P_{3n} \sigma_{3n+1}^- \sigma_{3n+2}^+ P_{3n+3} \Big).
\end{eqnarray}
The lowest weight state of $\bar{H}^z$ is $\vert \mathbb{Z}_3 \rangle$, as it should be, although it is degenerate. The first order perturbation will lift this degeneracy such that $\vert \mathbb{Z}_3 \rangle$ is the unique ground state of $\bar{H}^z_{(1)}$. We find the $\bar{H}^z, \bar{H}^+, \bar{H}^-$ obey the commutation relations:
\begin{eqnarray}
    [\bar{H}^z,\bar{H}^{+}] &=& \bar{H}^+ + \delta_{(1)}^+,  \\
\nonumber \delta_{(1)}^+ &=&  -\frac{1}{2}\sum_n \Big (  P_{3n-1}P_{3n} \sigma_{3n+1}^+ P_{3n+2} \\
\nonumber  &+& P_{3n-2} \sigma_{3n-1}^+ P_{3n} P_{3n+1}+ P_{3n-1} \sigma^-_{3n}P_{3n+1}P_{3n+2} \\
\nonumber  &+& P_{3n+1} P_{3n+2} \sigma_{3n+3}^- P_{3n+4} \Big ) \\
\nonumber 
&+& \frac{1}{2} \sum_n \Big ( P_{3n-1} \sigma_{3n}^- \sigma_{3n+1}^+ \sigma_{3n+2}^- P_{3n+3} \\
\nonumber  &+& P_{3n} \sigma_{3n+1}^- \sigma_{3n+2}^+ \sigma_{3n+3}^- P_{3n+4} \Big )\\
\nonumber  &+& \sum_n \Big ( P_{3n} \sigma^+_{3n+1} P_{3n+2} P_{3n+3} \\
&+& P_{3n} P_{3n+1} \sigma^+_{3n+2} P_{3n+3} \Big ).
\end{eqnarray}
Similarly, we find $[\bar{H}^z,\bar{H}^-] = - \bar{H}^- + \delta^-_{(1)}$, such that $\{\bar{H}^z,\bar{H}^+,\bar{H}^-\}$ form a broken representation of $\mathrm{su}(2)$. We identify the following first order perturbations to the PXP model which improve the representation:
\begin{eqnarray}
\nonumber    V_1 &=& \sum_n \Big(P_{3n-2} \sigma_{3n-1}^x P_{3n} P_{3n+1} + P_{3n-1} P_{3n} \sigma_{3n+1}^x P_{3n+2} \\
    &+& P_{3n-1} \sigma_{3n}^x P_{3n+1} P_{3n+2} + P_{3n-2} P_{3n-1} \sigma_{3n}^x P_{3n+1} \Big), \label{eq:z3_1} \\
\nonumber V_2 &=& \sum_n \Big(P_{3n} P_{3n+1} \sigma_{3n+2}^x P_{3n+3} 
+ P_{3n} \sigma_{3n+1}^x P_{3n+2} P_{3n+3} \Big), \\ && \label{eq:z3_2} \\
\nonumber V_3 &=& \sum_n \Big(P_{3n} \sigma_{3n+1}^x \sigma_{3n+2}^x \sigma_{3n+3}^x P_{3n+4} \\
    &+& P_{3n-1} \sigma_{3n}^x \sigma_{3n+1}^x \sigma_{3n+2}^x P_{3n+3} \Big).
    \label{eq:z3_3}
\end{eqnarray}
We emphasize that perturbations that improve $\mathbb{Z}_3$ revival, even at first order, break the full translation symmetry of the model to a subgroup of translations by a unit cell of size 3. This is different from $\mathbb{Z}_2$ revivals where the first-order corrections respect the full translation symmetry of the chain. We next discuss two interesting limits, corresponding to weak and strong magnitude of these perturbations.

\subsection{Weak limit}\label{sec:z3weak}

\begin{figure}
    \centering
    \includegraphics[width=0.5\textwidth]{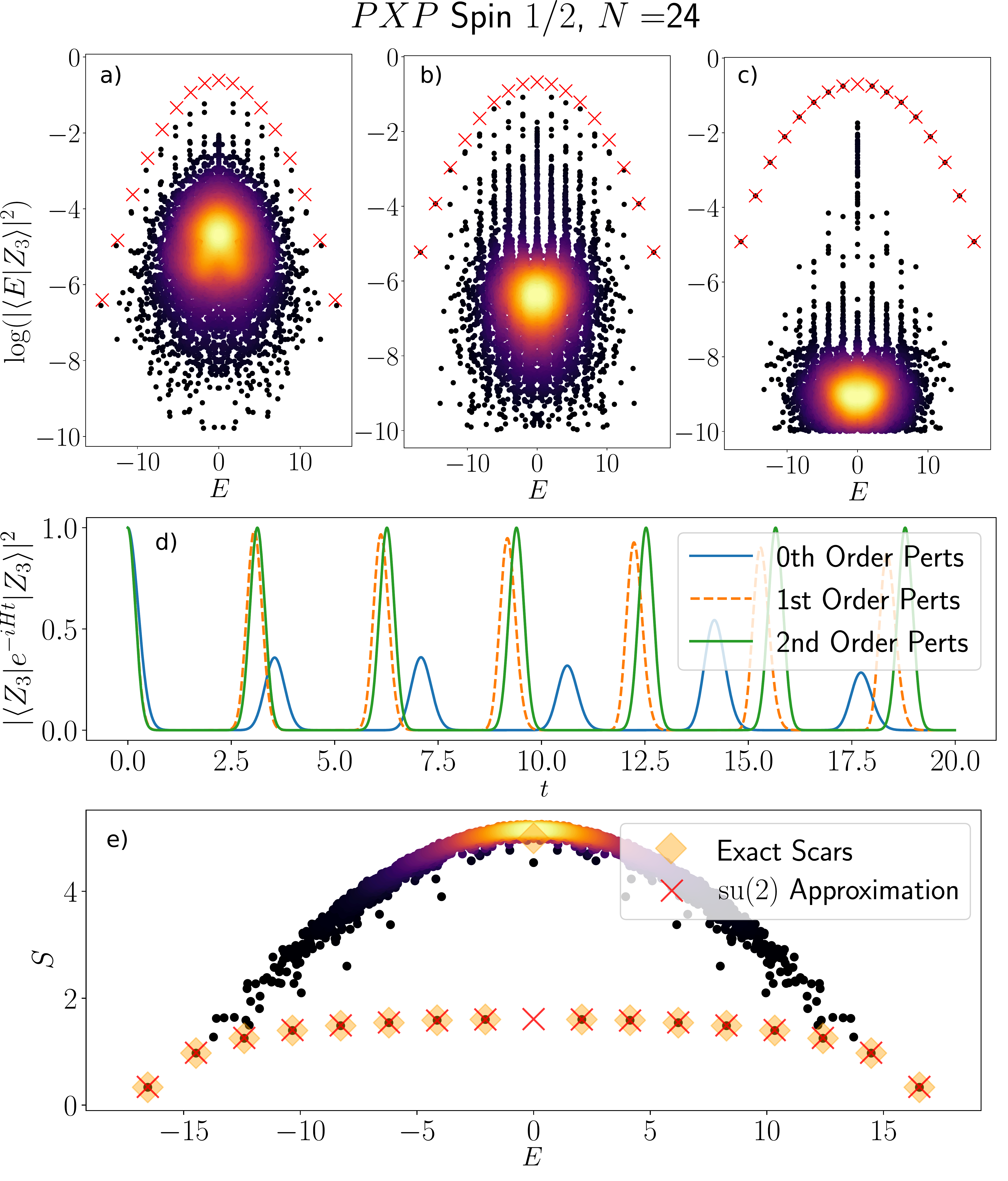}
    \caption{Improving the $\mathbb{Z}_3$ revival in the PXP model. (a) Eigenstate overlap with $\vert \mathbb{Z}_3 \rangle$ state for PXP model. (b) Eigenstate overlap after including first order correction in Eq.~(\ref{eq:z3_1})-(\ref{eq:z3_3}). (c) Eigenstate overlap after including second order perturbations listed in Appendix ~\ref{appendix:Z3_Perts}.  (d) Quantum fidelity when the system is quenched from $\vert \mathbb{Z}_3 \rangle$ state at various perturbation orders. The perturbation coefficients are those which maximize the first fidelity revival peak. (e) Bipartite entropy (Eq.~\ref{eq:entropy}) of eigenstates of the PXP model after including second order $\mathbb{Z}_3$ $\mathrm{su(2)}$ corrections. Points labelled ``Exact Scars" are exact diagonalization results identified from the top band of states in (c). Red crosses in (a), (b), (c), (e) indicate approximations to the scar states obtained by projecting the Hamiltonian to the broken representation basis and diagonalizing. Color scale in (a), (b), (c), (e) indicates the density of data points, with lighter regions being more dense.}
    \label{fig:z3PxpSpinHalf}
\end{figure}

By numerical optimization of the revival amplitude under perturbations in Eqs.~(\ref{eq:z3_1}), (\ref{eq:z3_2}) and (\ref{eq:z3_3}),  bounding coefficients to satisfy $\vert \lambda_i \vert <0.5$, we find that revivals from $\vert \mathbb{Z}_3 \rangle$ can be enhanced with optimal perturbation coefficients 
\begin{eqnarray}
\lambda^* = [0.18244,-0.10390,0.05445].
\end{eqnarray}
Similar to $\vert \mathbb{Z}_2 \rangle$ revival, we can find second order perturbations which improve revivals further (see Appendix~\ref{appendix:Z3_Perts} for the terms and optimal coefficients). A summary of the effect of succesive pertubations on $\vert \mathbb{Z}_3 \rangle$ is given in Fig.~\ref{fig:z3PxpSpinHalf}, while error metrics at various orders are given in Table~\ref{tab:pxpz3}.  Despite long-lived coherent oscillations when the system is initialized in the $|\mathbb{Z}_3\rangle$ state, we verify the model including second order perturbations is still ergodic by calculating the mean level spacing~\cite{OganesyanHuse} $\langle r \rangle = 0.5256$ at $N=24$, consistent with the Wigner-Dyson distribution one would expect in an ergodic system.
\begin{table}
    \centering
    \begin{tabular}{|c|c|c|c|c|}
        \hline
        Order & $1-f_0$ & $\sigma/D_{\mathrm{su(2)}}$ & $max(var(H^z)_n)$ & $K$ \\
        \hline
        $n=0$ & $6.397 {\times} 10^{-1}$ & $3.358 {\times} 10^{-1}$ & $9.300 {\times} 10^{-1}$ & $1.234 {\times} 10^{1}$ \\
        \hline
        $n=1$ & $1.338 {\times} 10^{-2}$ & $3.349 {\times} 10^{-2}$ & $1.717 {\times} 10^{-1}$ & $4.957 {\times} 10^{0}$ \\
        \hline
        $n=2$ & $1.852 {\times} 10^{-5}$ & $7.082 {\times} 10^{-3}$ & $2.357 {\times} 10^{-2}$ & $2.124 {\times} 10^{0}$\\
        \hline
    \end{tabular}
    \caption{Error metrics for the $\mathbb{Z}_3$ $\mathrm{su(2)}$ subspace of the PXP model at various perturbation orders for system size $N=24$. Subspace variance $\sigma$ is normalized by the dimension of the $\mathrm{su(2)}$ representation, $2N/3+1$. See text for details of the perturbations.}
    \label{tab:pxpz3}
\end{table}

\subsection{Strong limit: exact dynamical symmetry}\label{section:pxpZ3Exact}

A curious feature of $\mathbb{Z}_3$ revivals is that the $\mathrm{su}(2)$ algebra can be made exact for the model
\begin{eqnarray}\label{eq:z3spec}
    H = \sum_n \tilde{\sigma}_n^x - V_1,
\end{eqnarray}
which is the PXP model from which we subtracted the $V_1$ perturbation defined previously in Eq.~(\ref{eq:z3_1}). As the strength of $V_1$ is order unity, this model should not be called  a ``perturbation" to the PXP model. For the model in Eq.~(\ref{eq:z3spec}), the raising operator is
\begin{eqnarray}
 \nonumber    \bar{H}^+ &=& \sum_n \Big((\mathbb{I} - (P_{3n-2} + P_{3n+2})) \bar{\sigma}_{3n}^- \\
        &+& (\mathbb{I} - P_{3n-1}) \bar{\sigma}_{3n+1}^+ 
        + (\mathbb{I} - P_{3n+4}) \bar{\sigma}_{3n+2}^+ \Big ), 
\end{eqnarray}
and, as before, $\bar{H}^- = (\bar{H}^+)^{\dagger}$, $\bar{H}^z = \frac{1}{2} [\bar{H}^+,\bar{H}^-]$, $H = \bar{H}^+ + \bar{H}^-$.  By inspection, it is easy to see the projectors $(\mathbb{I}-P_{3n-1}), (\mathbb{I}-P_{3n+4})$ evaluate to zero when $\bar{H}^+$ is applied to $\vert \mathbb{Z}_3 \rangle = \vert 100100...\rangle$. Thus, the terms containing $\bar{\sigma}^+_{3n+1}, \bar{\sigma}^+_{3n+2}$ never generate a spin flip and spins pointing down at these sites are frozen. It follows that the action of $\bar{H}^+$ on $\vert \mathbb{Z}_3\rangle$ is equivalent to:
\begin{equation}
 (\bar{H}^+)^n \vert \mathbb{Z}_3 \rangle = \left( - \sum_n \tilde{\sigma}_{3n}^- \right)^n \vert \mathbb{Z}_3 \rangle, 
\end{equation}
which implies that, within this subspace, the $\mathrm{su}(2)$ algebra is exact. Dynamics is just a free precession of spins located at positions $3n$ along the chain, $\vert 100100...\rangle \rightarrow \vert 000000...\rangle \rightarrow \vert 100100..\rangle \rightarrow...$. The model now possesses an exact dynamical symmetry within the $\mathrm{su(2)}$ subspace, namely 
\begin{eqnarray}
    \big[P_{\mathrm{su(2)}}^{\dagger}HP_{\mathrm{su(2)}},P_{\mathrm{su(2)}}^{\dagger}Q^+ P_{\mathrm{su(2)}}\big] &=& P_{\mathrm{su(2)}}^{\dagger}Q^+ P_{\mathrm{su(2)}},\label{eq:z3ExactDynamicalSym} \\
    Q^+ = e^{-i \frac{\pi}{2} \bar{H}^y } \bar{H}^+ e^{i \frac{\pi}{2} \bar{H}^y}, \quad \bar{H}^y &=& \frac{1}{2i} (\bar{H}^+ - \bar{H}^-), \quad
\end{eqnarray}
where $P_{\mathrm{su(2)}}$ is the basis transformation which projects to the subspace spanned by the $\mathrm{su(2)}$ basis states $\vert n \rangle = (\bar{H}^+)^n \vert \mathbb{Z}_3 \rangle$.
 
The Hamiltonian in Eq.~(\ref{eq:z3spec}) fractures the Hilbert space in the computational basis even further than the pure PXP model. We find the number of sectors grows exponentially with system size, in a similar fashion to fractonic systems~\cite{Pretko2019}. While one sector is the desired embedded representation of $\mathrm{su(2)}$, various other sectors emerge due to the projectors blocking access from one configuration to another based on the decomposition of the state into unit cells of three consisting of $\{\vert 000 \rangle, \vert 001 \rangle, \vert 010 \rangle, \vert 100 \rangle, \vert 101 \rangle \}$. 

We find it is also possible for a model to feature an exactly embedded $\mathrm{su(2)}$ representation for which the computational basis does not fracture into exponentially many sectors as seen in the $\mathbb{Z}_3$ case. In the following Section we discuss one embedded $\mathrm{su(2)}$ representation which allows us to identify such a model.

 \section{$\mathbb{Z}_4$ Revivals from $\mathrm{su(2)}$ Algebra}\label{sec:pxpz4}

Unlike $\vert \mathbb{Z}_2 \rangle$ and $\vert \mathbb{Z}_3 \rangle$, quenches from $\vert \mathbb{Z}_4 \rangle = \vert 10001000...\rangle$ do not result in a reviving wavefunction beyond system size $N\gtrsim 20$  and expectation values of local observables equilibrate as expected from the ETH, such that there appear to be no scarred eigenstates with enhanced support on $\vert \mathbb{Z}_4 \rangle$.

 Nevertheless, in this Section we show that our Lie algebra approach identifies deformations to the PXP model which fixes a new $\mathrm{su}(2)$ algebra, engineered such that $\vert \mathbb{Z}_4 \rangle$ is the lowest weight eigenstate of some $\bar{H}^z$, rather than $\vert \mathbb{Z}_2\rangle, \vert \mathbb{Z}_3 \rangle$ as seen previously. While the subspace variance of this representation is too large to witness observable revivals in the PXP model, by fixing the algebra we realize new models which \emph{do} exhibit $\mathbb{Z}_4$ revivals.

In direct analogy with the previous cases, we define the raising and lowering operators as 
\begin{eqnarray}\label{eq:z4raising}
    \bar{H}^+ &=& \sum_n \left( \tilde{\sigma}^-_{4n} + \tilde{\sigma}^+_{4n+1} + \tilde{\sigma}^+_{4n+2} + \tilde{\sigma}_{4n+3}^+ \right), \\
    \bar{H}^- &=& \sum_n \left( \tilde{\sigma}^+_{4n} + \tilde{\sigma}^-_{4n+1} + \tilde{\sigma}^-_{4n+2} + \tilde{\sigma}_{4n+3}^- \right),
\end{eqnarray}
which, in turn, define $\bar{H}^z = \frac{1}{2} [\bar{H}^+,\bar{H}^-]$ that evaluates to 
\begin{eqnarray}
    \nonumber \bar{H}^z &=& \sum_n \left( -\tilde{\sigma}_{4n}^z + \tilde{\sigma}_{4n+1}^z + \tilde{\sigma}_{4n+2}^z + \tilde{\sigma}_{4n+3}^z \right) \\
\nonumber &+& \frac{1}{2} \sum_n \Big( P_{4n} \sigma_{4n+1}^+ \sigma_{4n+2}^- P_{4n+3} \\
\nonumber &+& P_{4n} \sigma_{4n+1}^- \sigma_{4n+2}^+ P_{4n+3} + P_{4n+1} \sigma_{4n+2}^+ \sigma_{4n+3}^- P_{4n+4} \\
 &+& P_{4n+1} \sigma_{4n+2}^- \sigma_{4n+3}^+ P_{4n+4} \Big). \quad
\end{eqnarray}
Similar to previous cases, $\vert \mathbb{Z}_4 \rangle$ is the lowest weight state of $\bar{H}^z$ and it is found that $\{ \bar{H}^z,\bar{H}^+,\bar{H}^-\}$ form a broken representation of $\mathrm{su}(2)$. Errors in the root structure (Appendix~\ref{appendix:pxpZ4_2ndOrder}) suggest the following perturbations to PXP model are necessary to stabilise $\mathbb{Z}_4$ revival:
\begin{eqnarray}
    V_1 &=& \sum_n P_{4n} \sigma^x_{4n+1} \sigma^x_{4n+2} \sigma^x_{4n+3} P_{4n+4}, \label{eq:z4Pert_1}\\
\label{eq:z4Pert_2}    \nonumber  V_2 &=& \sum_n \big(P_{4n-1} \sigma^x_{4n} \sigma^x_{4n+1} \sigma^x_{4n+2} P_{4n+3} \\
   &+& P_{4n+1} \sigma^x_{4n+2} \sigma^x_{4n+3} \sigma^x_{4n+4} P_{4n+5} \big), \\
\label{eq:z4Pert_3}   \nonumber      V_3 &=& \sum_n \big(P_{4n} P_{4n+1} \sigma^x_{4n+2} P_{4n+3} \\
\nonumber        &+& P_{4n} \sigma^x_{4n+1} P_{4n+2} P_{4n+3} \\
\nonumber        &+& P_{4n+1} P_{4n+2} \sigma^x_{4n+3} P_{4n+4} \\
\ &+& P_{4n+1} \sigma^x_{4n+2} P_{4n+3} P_{4n+4} \big), \\
      \nonumber   V_4 &=& \sum_n \big(P_{4n-2} \sigma^x_{4n-1} P_{4n} P_{4n+1} \label{eq:z4Pert_4} \\
\nonumber        &+& P_{4n-1} P_{4n} \sigma^x_{4n+1} P_{4n+2} \\
\nonumber        &+& P_{4n-1} \sigma^x_{4n} P_{4n+1} P_{4n+2}  \\
 &+& P_{4n+2} P_{4n+3} \sigma^x_{4n+4} P_{4n+5} \big). 
\end{eqnarray}

\begin{figure}
    \centering
    \includegraphics[width=0.5\textwidth]{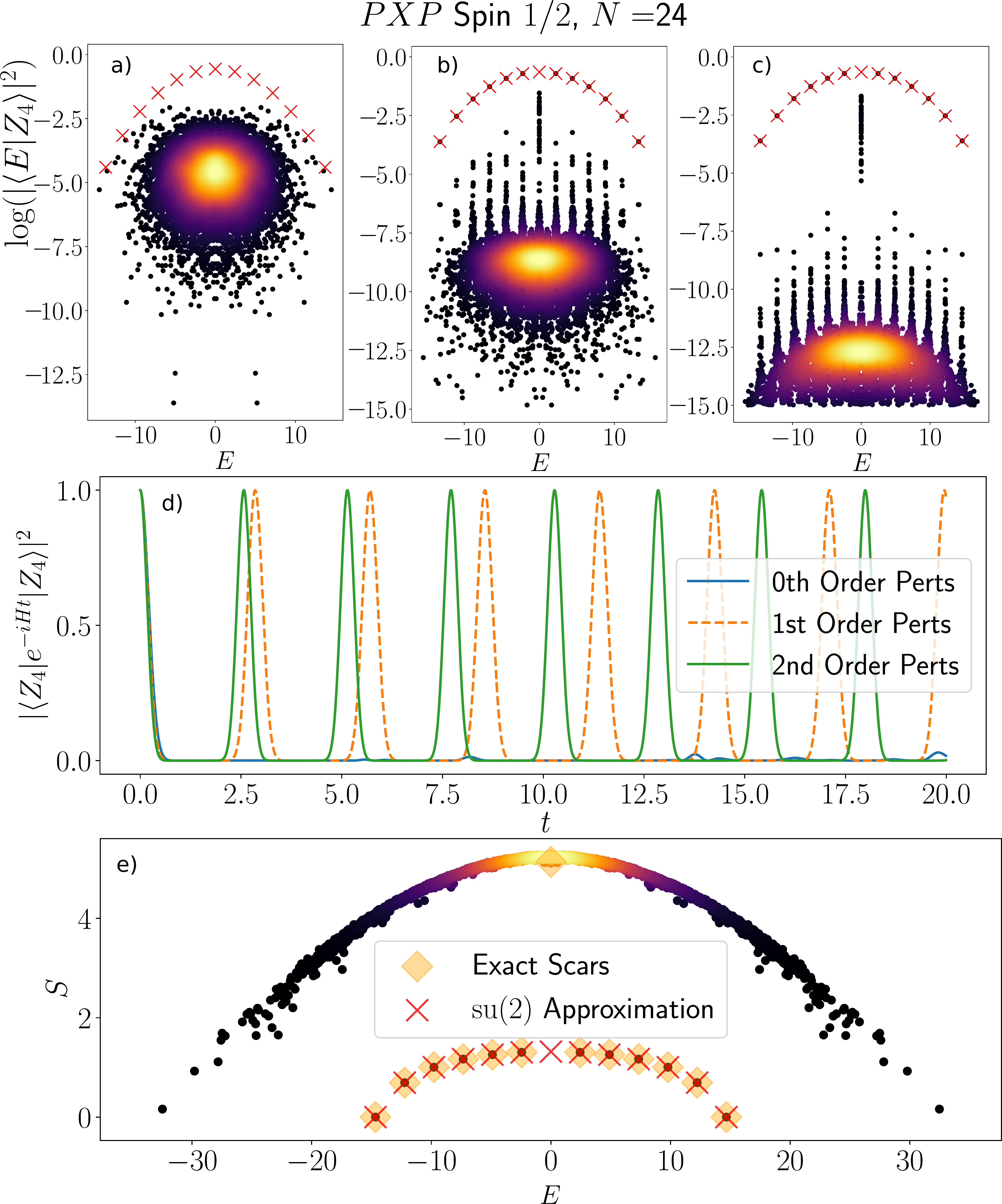}
    \caption{$\mathbb{Z}_4$ revival in PXP model. (a) Eigenstate overlap with $\vert \mathbb{Z}_4 \rangle$ state for PXP model. (b) Eigenstate overlap with $\vert \mathbb{Z}_4 \rangle$ state after including first order $\mathrm{su}(2)$ corrections, Eqs.~(\ref{eq:z4Pert_1})-(\ref{eq:z4Pert_2}). (c) Eigenstate overlap after including second order $\mathrm{su}(2)$ corrections (Appendix~\ref{appendix:pxpZ4_2ndOrder}).  (d) $\mathbb{Z}_4$ quench fidelity. $\vert \mathbb{Z}_4 \rangle$ state does not revive in pure PXP model, but it does revive in the new model obtained by correcting the $\mathrm{su}(2)$ algebra. (e) Bipartite entropy (Eq.~\ref{eq:entropy}) of eigenstates of the PXP model after including second order $\mathbb{Z}_4$ $\mathrm{su(2)}$ corrections. Points labelled ``Exact Scars" are exact diagonalization results identified from the top band of states in (c). Red crosses in (a), (b), (c), (e) indicate approximate scar states obtained by projecting the Hamiltonian to the broken representation basis and diagonalizing. Color scale in (a), (b), (c), (e) indicates the density of data points, with lighter regions being more dense.}
    \label{fig:pxpZ4summ}
\end{figure}
In contrast to our previous example of $\mathbb{Z}_3$ revival, explicit optimization finds that the terms in Eqs.~(\ref{eq:z4Pert_1})-(\ref{eq:z4Pert_4}) can stabilise $\mathbb{Z}_4$ revivals, but some of the resulting optimal coefficients turn out to be of the order unity. Thus, similar to the special $\mathbb{Z}_3$ case discussed above, we arrive at a model that cannot be viewed as a small deformation of PXP, but rather a new model in its own right. Specifically, optimizing $V_i$ coefficients $\lambda_i$ for fidelity we find (at $N=16$) 
\begin{eqnarray}
 \lambda_i^* = [0.0008,-1.43,0.0979,0.0980],
\end{eqnarray}
where we see the coefficient of optimal $V_2$ is $\sim O(1)$.  Once again, second order perturbations can be identified from the Lie algebra and revivals enhanced further (see Appendix~\ref{appendix:pxpZ4_2ndOrder} for details of the $36$ terms and optimal coefficients -- note only 3 terms contribute significantly with $O(1)$ coefficient  after optimizing for revivals). The effect of these perturbations is summarized in Fig.~\ref{fig:pxpZ4summ}. Error metrics at various perturbation orders are given in Table~\ref{tab:pxpZ4Errors}.  As in the previous examples, the second order deformations leave the model non-integrable, which we verify from the mean level spacing $\langle r \rangle = 0.5271$ at $N=24$, consistent with an ergodic system.
\begin{table}
    \begin{tabular}{|c|c|c|c|c|}
        \hline
        Order & $1-f_0$ & $\sigma/D_{\mathrm{su(2)}}$ & $max(var(H^z)_n)$ & $K$ \\
        \hline
        $n=0$ & $9.993 {\times} 10^{-1}$ & $3.333 {\times} 10^{0}$ & $2.779 {\times} 10^{0}$ & $4.323 {\times} 10^{0}$ \\
        \hline
        $n=1$ & $5.814 {\times} 10^{-5}$ & $6.722 {\times} 10^{-4}$ & $7.902 {\times} 10^{-4}$ & $3.258 {\times} 10^{-3}$ \\
        \hline
        $n=2$ & $3.351 {\times} 10^{-9}$ & $9.746 {\times} 10^{-6}$ & $2.753 {\times} 10^{-4}$ & $1.534 {\times} 10^{-3}$ \\
        \hline
    \end{tabular}
    \caption{Error metrics for the $\mathbb{Z}_4$ $\mathrm{su(2)}$ subspace of the PXP model at various perturbation orders for $N=24$. Subspace variance $\sigma$ is normalized by the dimension of the $\mathrm{su(2)}$ representation, $N/2+1$. See text for details of the perturbations. Errors at $n=0$ are much worse than $n=0$ $\mathbb{Z}_2, \mathbb{Z}_3$ errors (compare with Table~\ref{tab:pxpz2} and Table~\ref{tab:pxpz3}), consistent with there being no revivals or $\mathbb{Z}_4$ scars in pure PXP model.}
    \label{tab:pxpZ4Errors}
\end{table}

\subsection{Exact Z4 $\mathrm{su(2)}$ Embedding}
\label{section:pxpZ4}

Finally, we mention that similar to $\mathbb{Z}_3$ case, there exists a deformation of PXP such that $\vert \mathbb{Z}_4 \rangle$ is the lowest weight state of an \emph{exact} $\mathrm{su}(2)$ representation. That model is obtained by redefining the raising operator in Eq.~(\ref{eq:z4raising}) according to
\begin{eqnarray}
    \nonumber   \bar{H}^+ &\rightarrow& \bar{H}^+ - V_2\\
    \nonumber    &=& \bar{H}^+ - \sum_n \Big( P_{4n+3} \sigma_{4n+4}^- \sigma_{4n+5}^+ \sigma_{4n+6}^- P_{4n+7} \\
      &+& P_{4n+1} \sigma_{4n+2}^- \sigma_{4n+3}^+ \sigma_{4n+4}^- P_{4n+5} \Big),
\end{eqnarray}
which yields the Hamiltonian:
\begin{eqnarray}
  \nonumber  H &=& \sum_n P_{n-1}\sigma^x_{n}P_{n+1} \\
    \nonumber  &-& \sum_n \big( P_{4n+3}\sigma^x_{4n+4}\sigma^x_{4n+5}\sigma^x_{4n+6}P_{4n+7} \\
               &+& P_{4n+1} \sigma^x_{4n+2} \sigma^x_{4n+3} \sigma^x_{4n+4} P_{4n+5} \big).\label{eq:z4ExactModel}
\end{eqnarray}
Similar to the $\mathbb{Z}_3$ case, this model features an exact dynamical symmetry within the $\mathrm{su(2)}$ subspace, with the symmetry generator taking the same form as Eq.~(\ref{eq:z3ExactDynamicalSym}).
However, unlike the $\mathbb{Z}_3$ case, the computational basis which satisfies the Rydberg constraint does not fracture into exponentially many sectors. There still exists an exact Krylov subspace generated by repeated application of the Hamiltonian on $\vert \mathbb{Z}_4 \rangle$ which is block diagonal with respect to the orthogonal thermalizing subspace, such that this model exhibits type (b) scarring described in Fig~\ref{fig:scarMechs} and the Krylov subspace is an exact $\mathrm{su(2)}$ representation.  We verify the model is still thermalizing in the orthogonal subspace by verifying the mean level spacing $\langle r \rangle = 0.5365$ at $N=24$, consistent with level spacings obeying the Wigner-Dyson distribution as expected for an ergodic subspace. 

\begin{figure}[h]
    \centering
    \includegraphics[width=0.5\textwidth]{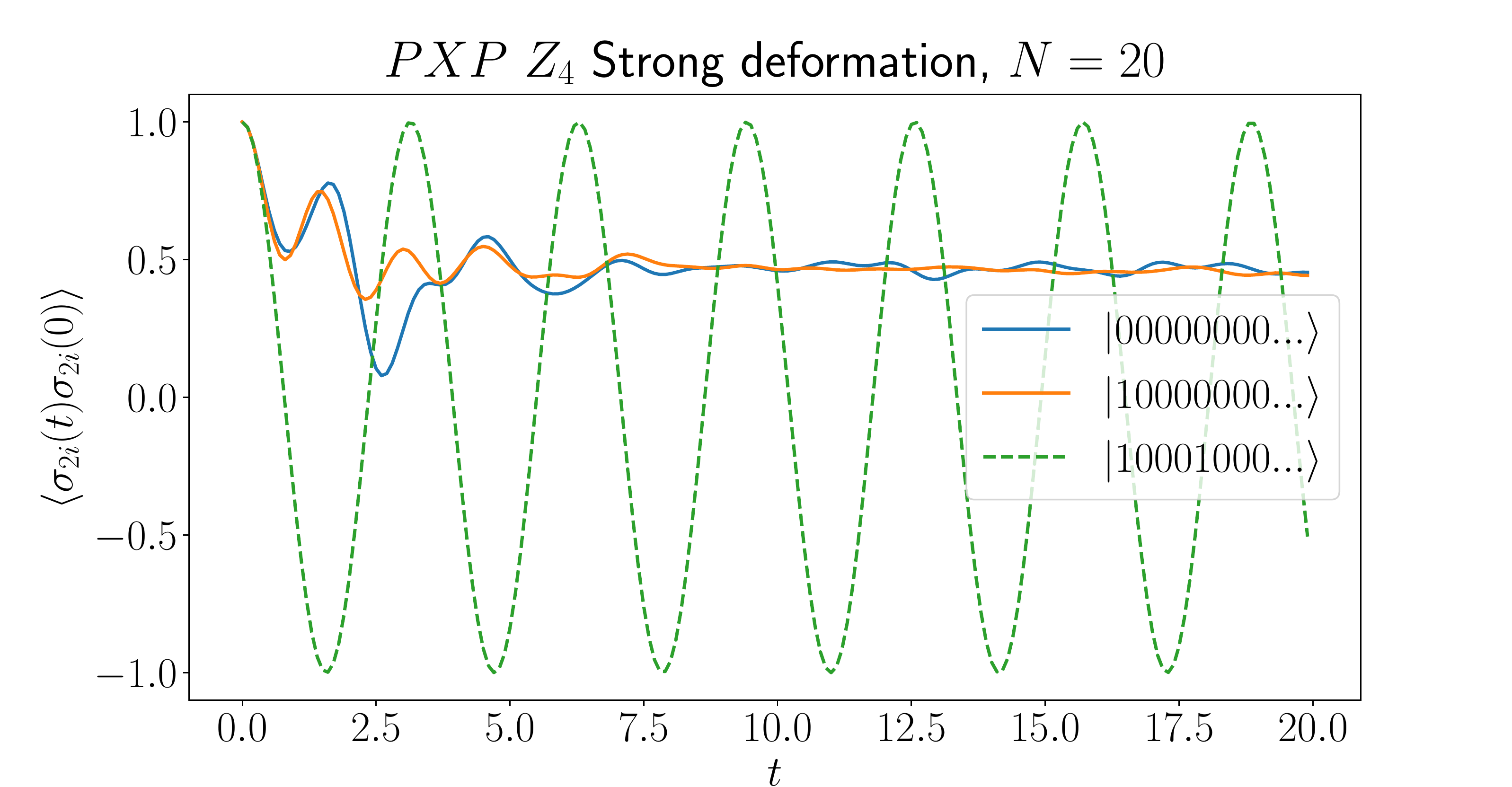}
    \caption{ Local autocorrelation function $\langle \sigma_{2i}^z(t) \sigma_{2i}^z(0) \rangle$ of the model given by Eq.~(\ref{eq:z4ExactModel}), for various initial states given in the legend. Results are for $N=20$. We consider sites $2i$ as the translation symmetry of Eq.~(\ref{eq:z4ExactModel}) is broken to a subgroup corresponding to translations by two units. Generic initial states such as the polarized state $\vert 000...\rangle$ equilibrate, whereas the autocorrelation function exhibits non stationary behaviour for all times when the system is initialized in the $\vert \mathbb{Z}_4 \rangle=\vert 10001000...\rangle$ state.
    }
    \label{fig:z4_autoc}
\end{figure}
As a consequence of the exact $\mathrm{su(2)}$ embedding the $\vert \mathbb{Z}_4 \rangle$ state revives perfectly, whereas generic initial states from the orthogonal sector still thermalize as expected from the ETH. Thus, local observables and local autocorrelation functions, which generically equilibrate, may exhibit long-lived non-stationary behavior following a quench from $\vert \mathbb{Z}_4 \rangle$,  Fig.~\ref{fig:z4_autoc}.

\section{Conclusions and Discussion}\label{sec:conc}

We have argued that, up to a rotation, many-body scars in kinetically constrained spin models can be interpreted as forming an approximate basis of a broken Lie algebra representation. This results in a loosely embedded integrable subspace with approximate dynamical symmetry, which acts as an approximate representation of the Lie algebra. Seeking deformations of the Hamiltonian which improve this broken Lie algebra we have identified several models related to the PXP model describing a chain of Rydberg atoms, which exhibit  many-body scars and feature near perfect revivals from the simple product states $\vert \mathbb{Z}_2 \rangle$, $\vert \mathbb{Z}_3 \rangle$, $\vert \mathbb{Z}_4 \rangle.$ Further, we have constructed two models with exactly embedded $\mathrm{su(2)}$ representations, thus obtaining ``exact scars" in a similar spirit to  ``Krylov-restricted thermalization"~\cite{MoudgalyaKrylov} and ``projector embedded" scar states~\cite{ShiraishiMori}. 

The identification of embedded $\mathrm{su(2)}$ subspaces followed from identifying decompositions of the Hamiltonian $H=\bar{H}^+ + \bar{H}^-$, with $\bar{H}^- = (\bar{H}^+)^{\dagger}$. Thus, the representation is fixed by the choice of $\bar{H}^+$. Obviously, this choice is not unique and many other possible decompositions of $H$ exist, but many of these decompositions would result in embedded representations whose subspace variance is too large to give rise to scarred dynamics.
However, from the examples considered above, it appears that aspects of an $\mathrm{su}(2)$ algebra can generically be improved in certain models like PXP, no matter how broken the representation is to begin with, by considering errors of a suitably defined broken representation (e.g. $ \mathbb{Z}_4$ case). An obvious question is ``how broken" can these representations be such that we would see signatures of $\mathrm{su}(2)$ dynamics (revivals) following quenches from states in the $\mathrm{su}(2)$ subspace. In the examples considered in the main text, subspace variance of the approximate representation basis seems to be the best indicator that one would see scarred dynamics. 

While the focus of this paper has been on deformations of the PXP model resulting in embedded $\mathrm{su(2)}$ representations, we note our construction can be readily applied to arbitrary spin chains.  An interesting question for future work is if it is possible to engineer approximate dynamical symmetries in a subspace without making use of a Lie algebra, but perhaps more general algebraic structures such as the quantum group $\mathrm{U_q(sl_2)}$. Indeed, exact dynamical symmetry of the Hamiltonian which does not rely on a Lie algebra root structure has already been observed in the AKLT model~\cite{BernevigEnt}. The model possesses a dynamical symmetry $[H_{\mathrm{AKLT}},K^+] = \omega K^+$ and, while the operators $\{K^+,K^- = (K^+)^{\dagger}, H^z=\frac{1}{2}[K^+,(K^+)^{\dagger}] \}$ form an exact representation of $\mathrm{su(2)}$, the AKLT Hamiltonian itself $H_{\mathrm{AKLT}}$ is not a linear combination of the $\mathrm{su(2)}$ generators. Therefore, the dynamical symmetry does not trivially follow from the root structure and further the scarred subspace, generated by repeated application of $K^{\pm}$ on the AKLT ground state, does not act as a representation of $\mathrm{su(2)}$~\cite{BernevigEnt}. Moreover, we have not considered embeddings of higher order $\mathrm{su(n)}$ Lie algebras throughout this paper, instead restricting only to $\mathrm{su(2)}$. We expect this to be increasingly more difficult compared to $\mathrm{su(2)}$, due to the presence of more than one set of raising operators, resulting in multiple error sources where there is no guarantee that improving errors of one set of raising operators will not exasperate errors in another set.

An important open question relates to the closure of the broken Lie algebra -- will recursively feeding higher order error terms back into the broken generators result in an exact representation? Indeed, we have identified two cases where an $\mathrm{su(2)}$ algebra can be made exact ($\vert \mathbb{Z}_3 \rangle, \vert \mathbb{Z}_4 \rangle$) after only considering first order error terms. Neglecting closure, we have demonstrated that this integrable subspace need not be exactly embedded, but can be loosely embedded with small enough subspace variance such that signatures of the embedded group are still realized in dynamics, as seen in the PXP model. Finally, it would be interesting to investigate generalizations of loosely embedded Lie algebras in the context of open quantum systems, where recent work has shown that dissipation can give rise to the emergence of kinetic constraints~\cite{Everest2016} and robust dynamical symmetry~\cite{Buca2019, Tindall2019}.

{\sl Note added:} During the completion of this manuscript we became aware of Ref.~\onlinecite{MotrunichTowers}, which clarifies further the ``exact scars" seen in models we describe in Section~\ref{section:IadecolaScars}.

\section{Acknowledgments}

We acknowledge support by EPSRC Grants No.  EP/R020612/1, No. EP/M50807X/1.
Statement of compliance with EPSRC policy framework on
research data: This publication is theoretical work that
does not require supporting research data. 
This research was supported in part by the National Science Foundation under Grants No. NSF PHY-1748958 and No. EP/R513258/1 (J.-Y.D).  We thank Berislav Bu\v ca and Gabriel Matos for their insightful comments on the manuscript.

\bibliography{references}

\appendix

    \section{Stabilizing $\mathrm{su(2)}$ algebra in spin-1 PXP model }
    
The work by Ho \emph{et al.}~\cite{wenwei18TDVPscar} pointed out that time-dependent variational principle (TDVP) can elegantly describe the $\mathbb{Z}_2$ revival in the PXP model if TDVP is applied to a manifold of matrix product states with low bond dimension, effectively resulting in a semiclassical description of scarred many-body dynamics. Furthermore, it was noticed that the same approach can be directly generalized to describe the spin-1 PXP model given by the
the same Hamiltonian as in Eq.~(\ref{eq:pxp}) where the flip and projector terms now act on three-level systems ($|0\rangle$, $|1\rangle$ and $|2\rangle$)  as follows:
\begin{eqnarray}
\sigma_n^x = \sqrt{2} \begin{pmatrix} 0 & 1 & 0 \\ 1 & 0 & 1 \\ 0 & 1 & 0 \end{pmatrix}, \quad P_n = \begin{pmatrix} 1 & 0 & 0 \\ 0 & 0 & 0 \\ 0 & 0 & 0 \end{pmatrix}.
\end{eqnarray}
As before, $\sigma^x$ is proportional to the standard spin-1 operator in $x$ direction and $P$ is the projector on the lowest weight state in $z$-direction (which we denote by $|0\rangle$). It has been established~\cite{Bull2019} that the spin-$1$ PXP model contains $2N+1$ scarred eigenstates with enhanced support on the N\'eel state $|\mathbb{Z}_2 \rangle \equiv \vert 0202...\rangle$. In this Appendix we demonstrate the scarred subspace of the spin-1 PXP model also acts as an approximate $\mathrm{su(2)}$ representation. Somewhat surprisingly, correcting the broken Lie algebra results in a different optimal perturbation as compared to the the spin-1 generalization of the spin-$1/2$ correction PPXP+PXPP (Eq.~\ref{eq:pxppert1}). 

\begin{figure}
    \centering
    \includegraphics[width=0.5\textwidth]{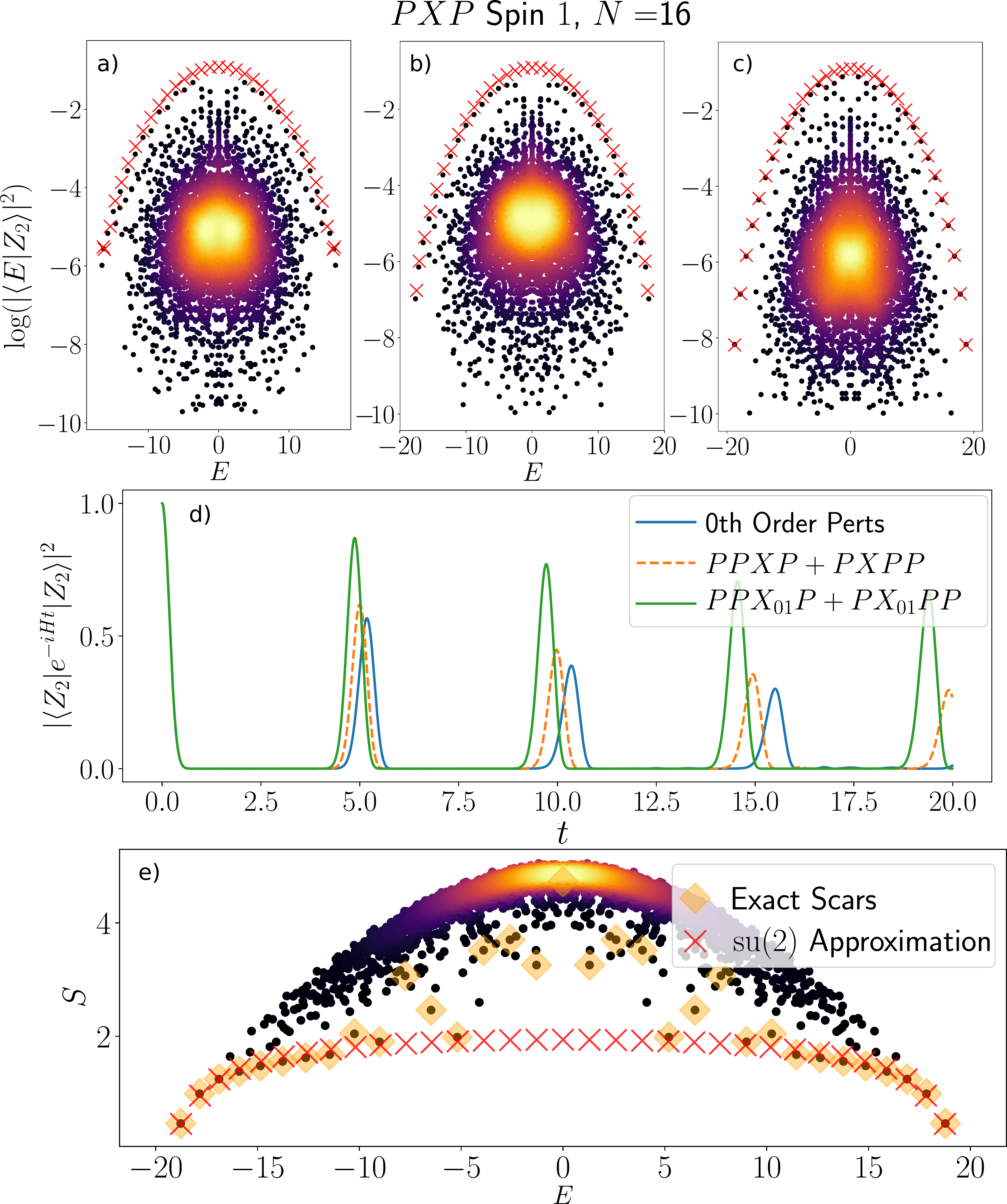}
    \caption{$\vert \mathbb{Z}_2 \rangle$ revival in PXP spin-$1$ model. (a) Eigenstate overlap with $\vert \mathbb{Z}_2 \rangle$ for pure PXP model. (b) Eigenstate overlap including the $PPXP+PXPP$ perturbation inspired by spin-$\frac{1}{2}$ PXP model. (c) Eigenstate overlap including first order $\mathrm{su(2)}$ correction, $PPX_{01}P+PX_{01}PP$. (d) $\vert \mathbb{Z}_2 \rangle$ quench fidelity with the various pertubations. e). Bipartite entropy of PXP spin-1 after including the first order $\mathbb{Z}_2$ $\mathrm{su(2)}$ correction $PPX_{01}P + PX_{01}PP$. Points labelled ``Exact Scars" are exact diagonalization results identified from the top band of states in c). Red crosses in (a), (b), (c), (e) indicate approximate scar states obtained by projecting the Hamiltonian to the broken representation basis and diagonalizing. Color scale in (a), (b), (c), (e) indicates the density of data points, with lighter regions being more dense.}
    \label{fig:spin1}
\end{figure}

We fix the broken $\mathrm{su}(2)$ representation by defining 
\begin{eqnarray}
    \bar{H}^+ &=& \sum_n \tilde{\sigma}^+_{2n} + \tilde{\sigma}^-_{2n-1},\\
    \bar{H}^z &=& \frac{1}{2}[\bar{H}^+,\bar{H}^-] = \sum_n \tilde{\sigma}^z_{2n} - \tilde{\sigma}^z_{2n-1}, 
\end{eqnarray}
using the same notation for $\tilde{\sigma}$ as in Eq.~(\ref{eq:sigmatilde}). The lowest weight state of $H^z$ is the N\'eel state, $\vert 0202... \rangle$. Checking the commutators, we arrive at the broken Lie algebra form
\begin{eqnarray}\label{eq:spin1}
    \nonumber [\bar{H}^z,\bar{H}^+] &=& \bar{H}^+ \\
      \nonumber       &-& \sqrt{2} \Big(PP(\sigma_{01})^+_{\mathrm{2n}}P + P(\sigma_{01})^+_{\mathrm{2n}}PP \\
            &+& P (\sigma_{01})^-_{\mathrm{2n+1}}PP + PP (\sigma_{01})^-_{\mathrm{2n+1}}P \Big), \\
          \nonumber  [\bar{H}^z,\bar{H}^{-}] &=& \bar{H}^- \\
        \nonumber  &+& \sqrt{2} \Big(PP(\sigma_{01})^-_{\mathrm{2n}}P + P(\sigma_{01})^-_{\mathrm{2n}}PP \\
            &+& P (\sigma_{01})^+_{\mathrm{2n+1}}PP + PP (\sigma_{01})^+_{\mathrm{2n+1}}P \Big), 
\end{eqnarray}
where we have introduced the operators
\begin{eqnarray}
\sigma_{01}^+ = \begin{pmatrix} 0 & 0 & 0 \\ 1 & 0 & 0 \\ 0 & 0 & 0 \end{pmatrix}, \quad \sigma_{01}^- = \begin{pmatrix} 0 & 1 & 0 \\ 0 & 0 & 0 \\ 0 & 0 & 0 \end{pmatrix},
\end{eqnarray} 
which we recognise as spin-$\frac{1}{2}$ raising and lowering operators. 
From Eq.~(\ref{eq:spin1}), we see that $\{\bar{H}^z,\bar{H}^+,\bar{H}^-\}$ form a broken representation of $\mathrm{su}(2)$. Lie algebra errors suggests the representation can be improved by perturbing $H$ with $V$:
\begin{eqnarray}
    \nonumber V &=& \sum_n P_{n-2}P_{n-1}\left( X_{01} \right)_n P_{n+1} \\
    &+& P_{n-1} \left( X_{01} \right)_n P_{n+1}P_{n+2},\;\;\;\;\;\; \\
     X_{01} &=& \begin{pmatrix} 0 & 1 & 0 \\ 1 & 0 & 0 \\ 0 & 0 & 0 \end{pmatrix}.
\end{eqnarray}
Importantly, we see that this perturbation is \emph{not} equal to PPXP+PXPP (the first-order correction term from the spin-$\frac{1}{2}$ PXP model).
Indeed, this perturbation is found to enhance revivals from the N\'eel state, with optimal coefficient $\lambda = 0.21423$ (at $N=12$). The revivals are significantly enhanced compared to the na\"ive perturbation ansatz PPXP+PXPP, with optimized coefficient $\lambda_{PPXP} = 0.05671$, see Fig.~\ref{fig:spin1} and a summary of error metrics in Table~\ref{table:spin1}.

\begin{table}
    \begin{tabular}{|c|c|c|c|c|}
        \hline
        & $1-f_0$ & $\sigma/D_{\mathrm{su(2)}}$ & $max(var(H^z)_n)$ & $K$ \\
        \hline
        $\textrm{No Pert}$ & 0.4330 & 0.4277 & 1.8711 & 22.3465\\
        \hline
        $PPXP$ & 0.3795 & 0.3118 & 1.3467 & 18.2751\\
        \hline
        $PPX_{01}P$ & 0.1304 & 0.0757 & 0.4124 & 14.1952\\
        \hline
    \end{tabular}
    \caption{Error metrics for $\mathbb{Z}_2$ revival in spin-1 PXP model after including two pertubations. $PPXP+PXPP$ is the pertubation one would expect to improve $\mathbb{Z}_2$ revivals based off $PXP$ spin $1/2$ results, whereas $PPX_{01}P+PX_{01}PP$ is the actual $\mathrm{su(2)}$ correction obtained from the broken root structure of the Lie algebra (see text for details). Results for $N=16$. Subspace variance $\sigma$ normalized by the dimension of the $\mathrm{su(2)}$ representation, $2N+1$. }
    \label{table:spin1}
\end{table}

\section{PXP $\mathbb{Z}_3$ second order $\mathrm{su(2)}$ perturbation terms}\label{appendix:Z3_Perts}

Here we detail the second order corrections to the embedded $\mathrm{su(2)}$ algebra which improves $\mathbb{Z}_3$ revivals, obtained by our recursive scheme summarized in Fig.~\ref{fig:algebraCorrection}. The second order perturbations to $H^{+}$, Eq.~(\ref{eq:z3raising}), are the following terms:
    \begin{eqnarray}
  \nonumber      \delta_{(1),0}^+ &=& P P \sigma^-_{3n} P + P \sigma^-_{3n} P P + P P \sigma^+_{3n+1} P + P \sigma^+_{3n+2} P P, \\
  \\
        \delta_{(2),1}^+ &=& PP \sigma_{3n}^- PP, \\
        \delta_{(2),2}^+ &=& P \sigma_{3n}^-P \sigma^z P + P \sigma^zP \sigma_{3n}^- P, \\
        \delta_{(2),3}^+ &=& P \sigma_{3n}^- P \sigma^z PP + PP \sigma^z P \sigma_{3n}^- P,\\
        \delta_{(2),4}^+ &=& P \sigma_{3n+1}^+ PP + PP \sigma_{3n+2}^+P, \\
        \delta_{(2),5}^+ &=& PP \sigma_{3n+1}^+ PP + PP \sigma_{3n+2}^+ PP, \\
        \delta_{(2),6}^+ &=& P \sigma_{3n+1}^+ PPP + PPP \sigma_{3n+2}^+ P, \\
        %\delta_{(2),7}^+ &=& P \sigma^z P \sigma_{3n+1}^+ P + P \sigma_{3n+2}^+ P \sigma^z P, \\
        \delta_{(2),7}^+ &=& P \sigma_{3n+1}^+ P \sigma^z P + P \sigma^z P \sigma_{3n+2}^+ P, \\
        \delta_{(2),8}^+ &=& P \sigma_{3n+1}^+ PQP + PQP \sigma_{3n+2}^+ P, \\
        \delta_{(2),9}^+ &=& PP \sigma_{3n+1}^+ PPP + PPP \sigma_{3n+2}^+ PP, \\
        \delta_{(2),10}^+ &=& PP \sigma_{3n+1}^+ P \sigma^z P + P \sigma^z P \sigma_{3n+2}^+ PP, \\
        \delta_{(2),11}^+ &=& P \sigma^+_{3n+1} P \sigma^z PP + PP \sigma^z P \sigma^+_{3n+2} P \\
        \delta_{(2),12}^+ &=& PP \sigma_{3n+1}^+ PQP + PQP \sigma_{3n+2}^+ PP, \\
        \delta_{(2),13}^+ &=& PP \sigma_{3n+1}^+ P \sigma^z PP + PP \sigma^z P \sigma_{3n+2}^+ PP, \\
        \delta_{(2),14}^+ &=& PPP \sigma_{3n+1}^+ P + P \sigma_{3n+2}^+ PPP,    
            \label{eq:appendixZ3_Perts}
    \end{eqnarray}
    where $Q\equiv \vert 1 \rangle \langle 1 \vert$. Perturbations to the PXP Hamiltonian follow from $V_{(n),m} = \delta_{(n),m}^+ + (\delta_{(n),m}^+)^{\dagger}$. Optimizing the coefficients of these terms at $N=18$, we find maximal wave-function revivals occur for:
        \begin{eqnarray}
\nonumber            \lambda_i^* &=& [0.1630,0.1129,0.0228,0.0409,\\
\nonumber                        &-&0.0871,0.0046,-0.0303,-0.0144,\\
\nonumber                        &-&0.0592,0.0005,0.0223,-0.0185,\\
                        &-&0.0451,0.0101,0.0035].
        \end{eqnarray}
    Thus, the dominant perturbations to the PXP Hamiltonian at second order are:
    \begin{eqnarray}
  \nonumber      V_1 &=& PP \sigma^x_{3n} P + P \sigma^x_{3n} P P + P P \sigma^x_{3n+1} P + P \sigma^x_{3n+2} P P, \\
  \\
        V_2 &=& PP \sigma^x_{3n} PP.    
    \end{eqnarray}

    \section{PXP $\mathbb{Z}_4$ second order $\mathrm{su(2)}$ pertubation terms}
    \label{appendix:pxpZ4_2ndOrder}
    
    For completeness, here we provide the full list of the $36$ second order corrections to the embedded $\mathrm{su(2)}$ representation responsible for $\mathbb{Z}_4$ revivals. These terms are identified by an iterative scheme summarized in Fig.~\ref{fig:algebraCorrection}. We do not consider every term which contributes an error to the broken Lie algebra but restrict to the subset of terms containing a single spin flip. Note, at second order only three perturbations to $H^+$ (Eq.~\ref{eq:z4raising}) dominate with coefficient $O(1)$ after optimizing for $\mathbb{Z}_4$ revivals. These are found to be:
    \begin{eqnarray}
        \delta_{1}^+ &=& P P Q P \sigma^+_{4n+3} P + P \sigma^+_{4n+1} P Q P P, \\
        \delta_{2}^+ &=& P P \sigma^+_{4n+2} P P, \\
        \delta_{3}^+ &=& P P \sigma^+_{4n+2} P\sigma^z PP + PP\sigma^z P \sigma_{4n+1} PP.
    \end{eqnarray}
    Optimizing the coefficients of all $36$ terms with respect to $\mathbb{Z}_4$ fidelity revivals at $N=16$ we find the coefficients of the above three terms are $[1.5621,1.9337,-1.4312]$. Before listing the full set of pertubations, we first introduce the following abbreviated notation:
    
    $$ ABC..., \quad m = \sum_i A_{4i+m} B_{4i+m+1} C_{4i+m+2}..., $$
    where $m$ is the offset of the far left operator from sites located at integer multiples of $4$.
    Listing multiple terms for a given perturbation is to be understood as implying addition with coefficient $1$.
    The complete set of second order $\mathbb{Z}_4$ $\mathrm{su(2)}$ corrections to $H^+$ are as follows: 
    \begin{eqnarray}
\nonumber           \delta_{(2),1}^+ &=& PP\sigma^{+}P, \quad 3 \\
\nonumber                           && P\sigma^{+}PP, \quad 2 \\
\nonumber                           && P\sigma^{-}PP, \quad 3 \\
                           && PP\sigma^{-}P, \quad 2 \\
\nonumber        \delta_{(2),2}^+ &=& PP\sigma^{+}PP, \quad 3 \\
                           && PP\sigma^{+}PP, \quad 1 \\
\nonumber        \delta_{(2),3}^+ &=& PPP\sigma^{+}P, \quad 3 \\
                           && P\sigma^{+}PPP, \quad 1 \\
\nonumber        \delta_{(2),4}^+ &=& PPQP\sigma^{+}P, \quad 3 \\
                           && P\sigma^{+}PQPP, \quad 0 \\
\nonumber        \delta_{(2),5}^+ &=& P\sigma^{-}PQP, \quad 3 \\
                           && PQP\sigma^{-}P, \quad 1 \\
\nonumber        \delta_{(2),6}^+ &=& P\sigma^-P\sigma^z P, \quad 3 \\
                           && P\sigma^z P\sigma^{-}P, \quad 1 \\
\nonumber        \delta_{(2),7}^+ &=& P\sigma^-P\sigma^z PP, \quad 3 \\
                           && PP\sigma^z P\sigma^{-}P, \quad 0 \\
\nonumber        \delta_{(2),8}^+ &=& P \sigma^z P\sigma^{+}P, \quad 3 \\
                           && P\sigma^{+}P\sigma^z P, \quad 1 \\
\nonumber        \delta_{(2),9}^+ &=& P\sigma^{+}PPP, \quad 2 \\
                           && PPP\sigma^{+}P, \quad 2 \\
\nonumber        \delta_{(2),10}^+ &=& PP\sigma^{+}PPP, \quad 1 \\
                            && PPP\sigma^{+}PP, \quad 2 \\
\nonumber        \delta_{(2),11}^+ &=& P\sigma^z P\sigma^{+}PP, \quad 3 \\
                            && PP\sigma^{+}P\sigma^z P, \quad 0 \\
        \delta_{(2),12}^+ &=& PP\sigma^{-}PP, \quad 2 \\
                 \nonumber   \delta_{(2),13}^+ &=& PP\sigma^{+}P, \quad 0 \\
         \nonumber                   && P\sigma^{+}PP, \quad 0 \\
         \nonumber                   &&PP\sigma^{+}P, \quad 1 \\
                              &&P\sigma^{+}PP, \quad 1 \\
        \delta_{(2),14}^+  &=& PP\sigma^{+}PP, \quad 0 \\
      \nonumber        \delta_{(2),15}^+  &=& P\sigma^{+}PPP, \quad 0 \\
                            &&PPP\sigma^{+}P, \quad 0 \\
    \nonumber       \delta_{(2),16}^+ &=& P\sigma^{+}PQP, \quad 0 \\
                            &&PQP\sigma^{+}P, \quad 0 \\
     \nonumber      \delta_{(2),17}^+ &=& PPQP\sigma^{-}P, \quad 0 \\
                            &&P\sigma^{-}PQPP, \quad 3 \\
     \nonumber      \delta_{(2),18}^+ &=& PP\sigma^{+}PPP, \quad 0 \\
                            &&PPP\sigma^{+}PP, \quad 3 
   \end{eqnarray}
 
    \begin{eqnarray}
    \nonumber       \delta_{(2),19}^+ &=& PQP\sigma^{+}P, \quad 3 \\
                            &&P\sigma^{+}PQP, \quad 1 \\
      \nonumber     \delta_{(2),20}^+ &=& PP\sigma^z P\sigma^{+}P, \quad 2 \\
                            &&P\sigma^{+}P\sigma^z PP, \quad 1 \\
       \nonumber    \delta_{(2),21}^+ &=& P\sigma^{-}PPP, \quad 3 \\
                            &&PPP\sigma^{-}P, \quad 1 \\
      \nonumber     \delta_{(2),22}^+ &=& P\sigma^z P\sigma^+P, \quad 0 \\
                            &&P\sigma^{+}P\sigma^z P, \quad 0 \\
     \nonumber      \delta_{(2),23}^+ &=& P\sigma^z P\sigma^{+}PP, \quad 0 \\
                            &&PP\sigma^{+}P\sigma^z P, \quad 3 \\
     \nonumber      \delta_{(2),24}^+ &=& PP\sigma^{-}PPP, \quad 2 \\
                            &&PPP\sigma^{-}PP, \quad 1 \\
     \nonumber      \delta_{(2),25}^+ &=& P\sigma^{+}P\sigma^z PP, \quad 0 \\
        \nonumber                       &&PP\sigma^z P\sigma^{+}P, \quad 1 \\
      \nonumber                         &&PP\sigma^z P\sigma^{+}P, \quad 3 \\
                            &&P\sigma^{+}P\sigma^z PP, \quad 2 \\
      \nonumber     \delta_{(2),26}^+ &=& P\sigma^z P\sigma^{+}P, \quad 2 \\
                            &&P\sigma^{+}P\sigma^z P, \quad 2 \\
      \nonumber     \delta_{(2),27}^+ &=& PP\sigma^{+}P\sigma^z P, \quad 1 \\
                            &&P\sigma^z P\sigma^{+}PP, \quad 2 \\
      \nonumber     \delta_{(2),28}^+ &=& PPP\sigma^{+}PP, \quad 0 \\
                            &&PP\sigma^{+}PPP, \quad 3 \\
       \nonumber    \delta_{(2),29}^+ &=& PP\sigma^{-}P\sigma^z P, \quad 2 \\
                            &&P\sigma^z P\sigma^{-}PP, \quad 1 \\
\nonumber            \delta_{(2),30}^+ &=& P\sigma^{+}PPPP, \quad 0 \\
                            && PPPP\sigma^{+}P, \quad 3 \\
\nonumber        \delta_{(2),31}^+ &=& PP\sigma^z P\sigma^{-}PP, \quad 0 \\
                            && PP\sigma^{-}P\sigma^z PP, \quad 2 \\
 \nonumber       \delta_{(2),32}^+  &=& PP\sigma^z P\sigma^{+}PP, \quad 1 \\
                            && PP\sigma^{+}P\sigma^z PP, \quad 1 \\
\nonumber        \delta_{(2),33}^+ &=& P\sigma^{+}PPPP, \quad 2 \\
                            && PPPP\sigma^{+}P, \quad 1 \\
\nonumber        \delta_{(2),34}^+ &=& PP\sigma^z P\sigma^{+}PP, \quad 3 \\
                            && PP\sigma^{+}P\sigma^z PP, \quad 3 \\
\nonumber        \delta_{(2),35}^+ &=& P\sigma^{+}PPPP, \quad 1 \\
                            && PPPP\sigma^{+}P, \quad 2 \\
\nonumber        \delta_{(2),36}^+ &=& PP\sigma^{+}P\sigma^z PP, \quad 0 \\
                            && PP\sigma^z P\sigma^{+}PP, \quad 2 
    \end{eqnarray}
Perturbations to the PXP Hamiltonian (Eq.~\ref{eq:pxp}) follow from $V_{(2),m} = \delta_{(2),m}^+ + (\delta_{(2),m}^+)^{\dagger}$. Optimizing coefficients of these terms at $N=16$ with respect to the first maximum of $\vert \langle \mathbb{Z}_4 \vert e^{-iHt} \vert \mathbb{Z}_4 \rangle \vert^2$ at $N=16$ we find:
\begin{eqnarray}
\nonumber    \lambda_i^* &=& [0.0888,0.2559,0.0796,1.5621,\\
\nonumber                && 0.1776,-0.0028,-0.0325,0.0099,\\
\nonumber                && 0.1333,0.0321,-0.0148,0.1490,\\
\nonumber                && 0.0728,1.9337,0.0001,0.0587,\\
\nonumber                && 0.0902,0.0001,0.1109,0.0104,\\
\nonumber                && 0.0468,0.0277,-0.0023,0.1046,\\
\nonumber                && 0.0667,0.0299,0.0437,0,\\
\nonumber                 && 0.0031,0.0002,-0.0189,0.0995,\\
                && 0.1531,0.0001,0.0001,-1.4312].
\end{eqnarray}

\end{document}